\documentclass[useAMS,usenatbib]{mn2e}

\usepackage{graphicx}
\usepackage{setspace}
\usepackage{natbib}
\usepackage{color}
\usepackage{amsmath,amssymb}
\usepackage{times}
\usepackage{aas_macros}

\title[Precession of the Sagittarius stream]{Precession of the
  Sagittarius stream}

\author[Belokurov et al.]{V. Belokurov$^{1}$\thanks{E-mail:vasily@ast.cam.ac.uk}, S. E. Koposov$^{1,2}$, N. W. Evans$^{1}$, J. Pe\~{n}arrubia$^{1,3}$, M. J. Irwin$^1$, M. C. Smith$^4$, \newauthor G. F. Lewis$^{1,5}$, M. Gieles$^{1,6}$, M. I. Wilkinson$^{7}$, G. Gilmore$^{1}$, E. W. Olszewski$^8$, \newauthor and M. Niederste-Ostholt$^{1}$
\\ $^{1}$Institute of Astronomy, Madingley Rd, Cambridge, CB3 0HA,
\\ $^{2}$Sternberg Astronomical Institute, Moscow State University,
Universitetskiy pr. 13, Moscow 119991, Russia\\
$^{3}$Institute for Astronomy, Royal Observatory, Blackford Hill View  Edinburgh, City of Edinburgh EH9 3HJ\\
$^{4}$Shanghai Astronomical Observatory, 80 Nandan Road, Shanghai 200030, China\\
$^{5}$Sydney Institute for Astronomy, School of Physics, A28, The University of Sydney, NSW 2006, Australia\\
$^{6}$Department of Physics, University of Surrey, Guildford, GU2 7XH, UK\\
$^{7}$Department of Physics \& Astronomy, University of Leicester, Leicester LE1 7RH\\
$^{8}$Steward Observatory, University of Arizona, Tucson, AZ 85721,USA
}

\begin{document}

\date{December 2012}
\pagerange{\pageref{firstpage}--\pageref{lastpage}} \pubyear{2012}

\maketitle

\label{firstpage}

\begin{abstract}
Using a variety of stellar tracers -- blue horizontal branch stars,
main-sequence turn-off stars and red giants -- we follow the path of
the Sagittarius (Sgr) stream across the sky in Sloan Digital Sky
Survey data. Our study presents new Sgr debris detections, accurate
distances and line-of-sight velocities that together help to shed new
light on the puzzle of the Sgr tails. For both the leading and the
trailing tail, we trace the points of their maximal extent, or
apo-centric distances, and find that they lie at $R^{\rm L} = 47.8 \pm
0.5$ kpc and $R^{\rm T} = 102.5 \pm 2.5$ kpc respectively. The angular
difference between the apo-centres is $93 \fdg 2 \pm 3 \fdg 5$, which
is smaller than predicted for logarithmic haloes. Such differential
orbital precession can be made consistent with models of the Milky Way
in which the dark matter density falls more quickly with
radius. However, currently, no existing Sgr disruption simulation can
explain the entirety of the observational data.  Based on its position
and radial velocity, we show that the unusually large globular cluster
NGC 2419 can be associated with the Sgr trailing stream. We measure
the precession of the orbital plane of the Sgr debris in the Milky Way
potential and show that, surprisingly, Sgr debris in the primary
(brighter) tails evolves differently to the secondary (fainter) tails,
both in the North and the South.
\end{abstract}


\section{Introduction}

\begin{figure*}
  \centering
  \includegraphics[width=17cm]{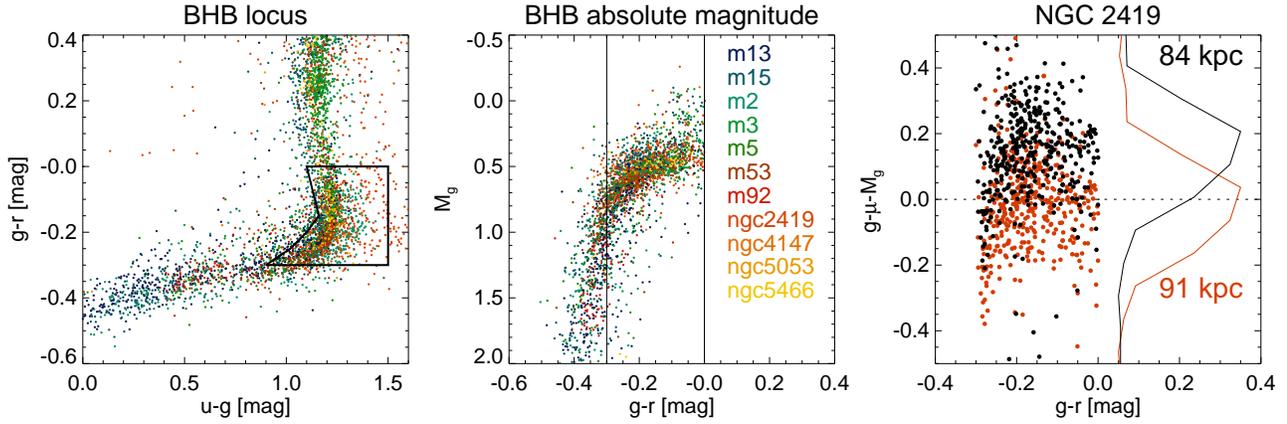}
  \caption[]{\small BHB stars in SDSS bands. \textit{Left:} BHB $u-g,
    g-r$ locus as traced by the BHB stars in the Galactic star
    clusters. Each dot represents a BHB candidate star coloured
    according to the membership in one of the 11 clusters listed in
    the middle panel. The polygon shows the region proposed byg
    \citet{Ya00} to select BHBs. \textit{Middle:} BHB absolute
    magnitude as a function of $g-r$ colour. Note how stars belonging
    to different star clusters group densely around similar intrinsic
    luminosity which changes slowly with $g-r$ colour. Vertical black
    lines mark the boundaries of the selection region shown in the
    Left panel. \textit{Right:} Offset between the distance modulus
    calculated using eq. 7 of \citet{De11} and the distance
    modulus suggested by \citet{Ha96} for the BHB candidate stars in
    the globular cluster NGC 2419. Red dots are for the assumed
    heliocentric distance to NGC 2419 of 91 kpc. Curves show the
    histograms of the distance modulus offset.}
   \label{fig:bhb}
\end{figure*}

As long as Newton's law of attraction holds true, an orbit gives the
most straightforward method of inferring the underling gravitational
potential. Such an inference is truly unambiguous when the orbit
mapped is nearly complete, as demonstrated beautifully by the recent
measurements of the mass of the Milky Way's central black hole
(e.g. \citealt{Gh08}). At larger Galactocentric distances, orbital
periods quickly grow to a significant fraction of the Hubble time,
rendering the tracking of a trajectory impractical. Yet, it is still
possible to establish the paths of some infalling Galactic fragments
rather accurately even at distances beyond many tens of kpc. Certain
Milky Way satellites sprout stellar tidal tails long enough to
delineate some of their orbit. These include the Sagittarius dSph
stream \citep{Ma03}, the tails of the Palomar 5 globular cluster
\citep{Od03}, as well as the Orphan \citep{Be07} and the GD-1
\citep{GD06} streams.

So far, the most robust inference of the Galactic potential based on
the measurements of a stellar stream has been the one carried out with
the GD-1 system by \citet{Ko10}. Even though only $\sim 100^{\circ}$
of arc of the stream has been mapped, the detailed phase-space information
and the intrinsic coldness of the stellar debris allowed the authors
to avoid the degeneracies inherent in the stream modelling process
(e.g. \citealt{Ey09}). However, given the orbit of the GD-1 stream,
the circular velocity measurement it facilitates, while independent
and robust, is limited to the range of galactocentric distances
accessible to other techniques (e.g. \citealt{Bo12}).

To reach into the outer Galactic halo, more distant and luminous
streams like the Orphan or Sagittarius are needed. The kinematically
cold Orphan stream seems to possess the properties ideal for orbit
inference. However, compared to the total of only $\sim 60^{\circ}$ of
the available detections of the Orphan Stream at distances between 30
and 50 kpc, there exists much more abundant information for both the
leading and the trailing tail of the Sgr stream. Each extends some
$\sim 180^{\circ}$ of arc, probing Galactocentric distances in the
range of 20-100 kpc. As of today, the narrow, but relatively faint
Orphan stream remains largely unstudied, yet a broad-brush picture
of the wide and luminous Sgr stream is slowly taking shape, thanks to
the diversity of the disruption models at hand (e.g. \citealt{Ib01,
  He04, Jo05, Fe06, LM10, Pe10}).

The most comprehensive model to date \citep{LM10} fits accurately the
3D positions and the radial velocities of the brightest components of
the leading and the trailing tail as mapped by \citet{Be06, La04}, and
\citet{Ma03, Ma04} respectively. This N-body simulation, however, does
not provide an explanation for the tail ``bifurcations'' observed both
in the North and the South \citep{Be06, Ko12}. Furthermore, the model
does not produce Sgr tidal debris running with the opposite distance
gradient at the locations of the leading tail around the North
Galactic Cap (Branch C in the notation of \citet{Fe06} as measured in
\citet{Be06}), or the distant the debris as detected by \citet{Ne03}
and recently confirmed by \citet{Ru11} in the direction of the
Galactic anti-centre. Naturally, given the wealth of the data
available for the Sgr system and the inherent complexity of the Milky
Way's gravitational potential, it is prudent to adopt the modelling
strategy in which some of the many unknowns are fixed at their
best-guess values. For their experiment, \citet{LM10} chose to fix
most of the orbital properties of the Sgr progenitor, as well as many
of the ingredients describing the Galactic potential. For example, the
potential of the dark matter halo is set to be logarithmic, while the
importance of changing the shape of the dark halo is explored.

In this Paper, based on the Sloan Digital Sky Survey Data Release 8
(SDSS DR8) we i) present several new detections of the Sgr debris in
the Galactic North, ii) refine some of the stream distances and iii)
substantiate the 3D data with the measurements of the stream's radial
velocity. As a result, we show that the Branch C \citep{Be06,Fe06} and
the anti-centre debris \citep{Ne03, Ru11} are all part of the very
long trailing tail of the Sagittarius dwarf. We show that at the very
edges of the SDSS footprint, both the leading and trailing tail reach
their respective apo-centres, thus indicating that the galactocentric
orbital precession (or apsidal precession) of the Sgr dSph is close to
$\sim 95^{\circ}$. This is to be compared with logarithmic haloes,
where the orbital precession is typically $\sim 120^{\circ}$. The
lower rate of the orbital precession signifies a sharper drop in the
dark matter density as a function of Galactocentric radius.  We also
measure the precession of the plane of the Sgr debris. This is related
to the Sgr orbital plane evolution under the torques from the disk and
the dark halo of the Milky Way.

\begin{figure*}
  \centering
  \includegraphics[width=0.98\linewidth]{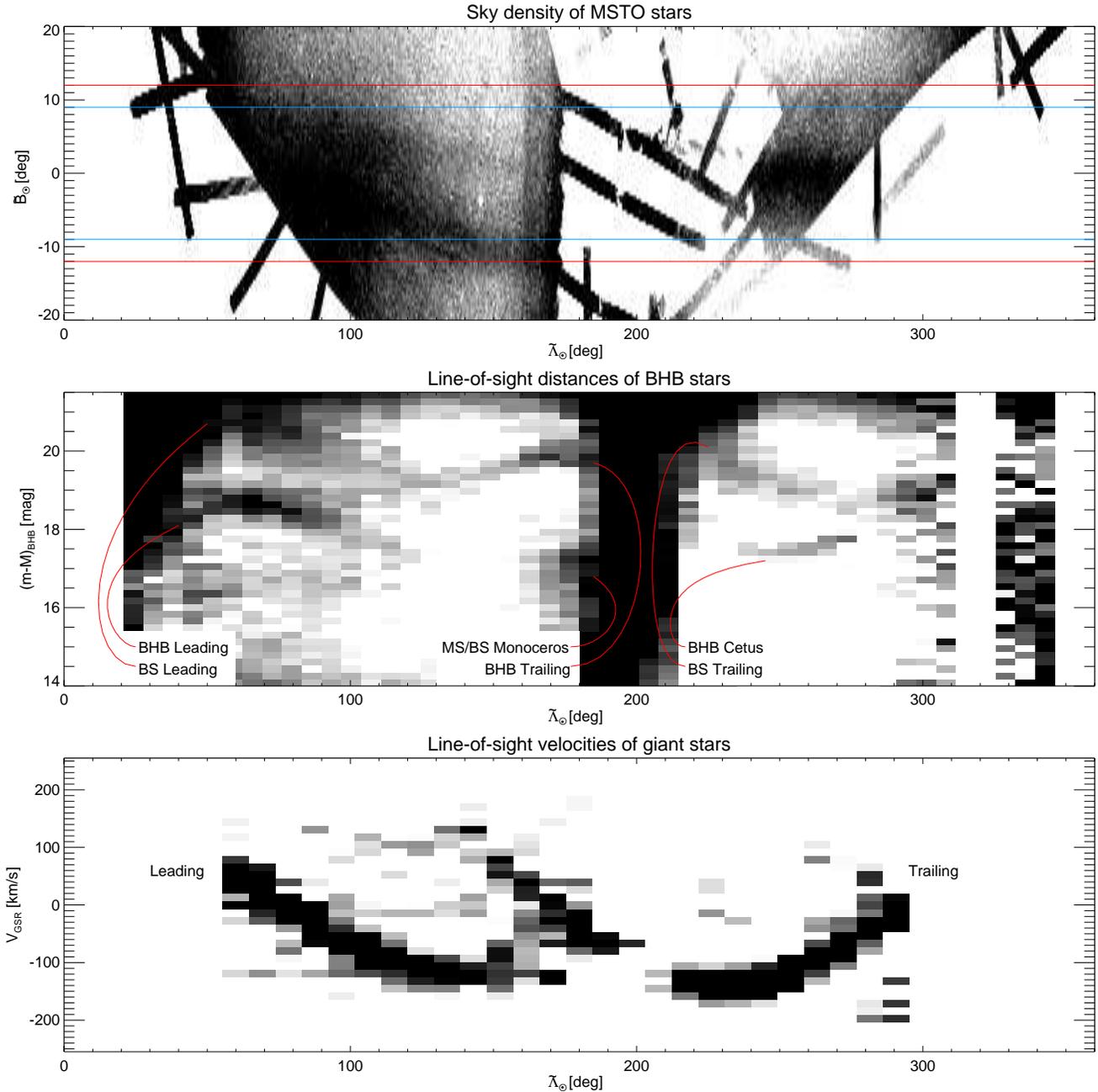}
  \caption[]{\small Sagittarius stream tomography with multiple
    tracers.  Darker regions correspond to enhanced stellar
    density. \textit{Top:} Sky density of MSTO stars in the Sgr stream
    coordinate system similar to that defined by \citet{Ma03}. Red
    (blue) lines show the range of latitude $B$ used to select stars
    for Middle (Bottom) panel. \textit{Middle:} Density of stream
    stars in the plane of Sgr stream longitude
    $\tilde{\Lambda}_{\odot}$ and distance modulus. For this plot, BHB
    candidate stars with $ -12^{\circ} < \tilde{B}_{\odot} <
    12^{\circ}$ are selected using the criteria of \citet{Ya00}, while
    the distances are assigned according to eq. 7 of \citet{De11}. In
    the North, the Sgr leading tail is clearly seen at $40^{\circ} <
    \tilde{\Lambda}_{\odot} < 120^{\circ}$ in both BHBs ($17< m-M
    <19$) and BSs ($19 < m-M < 21.5$). Note also the unambiguous
    detection of the trailing debris at $130^{\circ} <
    \tilde{\Lambda}_{\odot} < 190^{\circ}$. In the South, the Sgr
      trailing debris can be traced with BS stars in the range
      $220^{\circ} < \tilde{\Lambda}_{\odot} <
      290^{\circ}$. \textit{Bottom:} Density of the stream stars in
      the plane of $\tilde{\Lambda}_{\odot}$ and the line-of-sight
      velocity $V_{\rm GSR}$. For this plot, giant stars with $
      -9^{\circ} < \tilde{B}_{\odot} < 9^{\circ}$ are selected using
      the criteria in eq. 1 from the SDSS spectroscopic
      database. Note, that the tentative trailing velocity signal at
      $140^{\circ} < \tilde{\Lambda}_{\odot} < 190^{\circ}$ appears to
      be the natural continuation of the radial velocity run in the
      South, only interrupted by the disk at $190^{\circ} <
      \tilde{\Lambda}_{\odot} < 210^{\circ}$}
   \label{fig:data}
\end{figure*}

\section{Stream tomography with multiple tracers}
\label{sec:data}

For all the subsequent analysis, we correct the SDSS photometry for
the effects of extinction using the dust maps of \citet{SFD}. To
correct for the solar reflex motion, we adopt $V_{\rm LSR}=235$ km/s
and $(U,V,W)=(11.1, 12.2, 7.25)$ \citep{Sc10}.

\subsection{Blue Horizontal Branch stars in SDSS and the distance to NGC 2419}

To measure mean heliocentric distances to portions of the Sgr stream
within the SDSS field of view, we use Blue Horizontal Branch (BHB)
stars. The luminosity of BHBs does not vary to any significant degree
with changing age and metallicity and is primarily a function of the
star's temperature, which makes them attractive standard
candles. \citet{Ya00} have shown that by using the SDSS $u-g$ and
$g-r$ colours, highly complete (for a particular range of effective
temperatures) samples of BHBs can be selected with low levels of
contamination. At high Galactic latitudes, the principal source of
false positives within the boundaries of the colour-colour box
proposed by \citet{Ya00} are the intrinsically less luminous Blue
Straggler (BS) stars. Note that the photometrically-selected BHB
sample analysed below has a non-zero BS contamination. This, however,
does not affect the results as we do not study individual A-coloured
stars, but instead measure the properties of stellar over-densities.

\begin{figure*}
  \centering
  \includegraphics[width=0.98\linewidth]{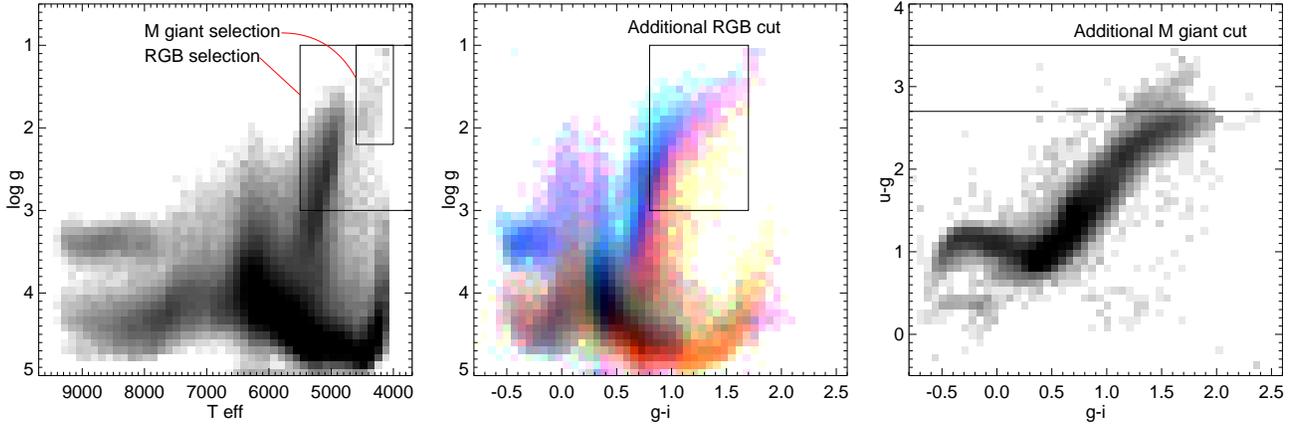}
  \caption[]{\small Selection of RGB and M giant stars from the SDSS
    spectroscopic sample. \textit{Left:} Density of stars with spectra
    in the SDSS DR8 in the plane of effective temperature $T_{eff}$
    and surface gravity $\log g$. Darker shades of grey mean higher
    density of stars. This is an analog of the familiar
    Hertzsprung-Russell diagram with the main sequence, the red giant
    branch and the blue horizontal branch all clearly visible. Solid
    lines give the boundaries used to select RGB and M giant candidate
    members of the stream. \textit{Middle:} Density of stars in color
    $g-i$ and surface gravity $\log g$ plane, color-coded according to
    metallicity $[Fe/H]$. Red, Green and Blue channels of the image
    are constructed with grey-scale density distributions of stars in
    the three metallicity bins: $-0.75 < [Fe/H] < 0$, $-1.5< [Fe/H] <
    -0.75$ and $-3 < [Fe/H] < -1.5$. The additional $g-i$ cut
    preferentially selects more metal-rich RGB stars. \textit{Right:}
    Density of the SDSS DR8 stars with spectra in the color-color
    plane of $g-i, u-g$. The brighter and/or the more metal-rich M
    giants stand out clearly thanks to their redder $u-g$ color. With
    this in mind, an additional $u-g$ cut is used to produce a cleaner
    selection of the Sgr M giant candidates.}
   \label{fig:tracers}
\end{figure*}

Figure \ref{fig:bhb} shows the properties of the candidate BHB stars
in 11 Galactic star clusters analysed by \citet{An08}. The left panel
of the Figure confirms the effectiveness of the colour cuts suggested
by \citet{Ya00}. BHBs from clusters with different stellar populations
essentially lie on top of each other forming a tight locus going
bluewards in $g-r$ at constant $u-g$ and eventually turning blue in
$u-g$ at around $g-r\sim -0.35$. As is obvious from the middle panel of
this Figure, the absolute magnitude of a BHB varies with its
$g-r$ color, as governed by changing temperature. At small negative
values of $g-r$, $M_g \sim 0.5$. However, on moving bluewards, the
Horizontal Branch tips and by $g-r=-0.35$, the absolute magnitude is at
least half a magnitude fainter. The behaviour of the centroid of the
distribution of the Galactic globular cluster BHBs in the plane of
absolute magnitude and $g-r$ colour can be described accurately with a
4th degree polynomial, as proposed by \citet{De11}. In what follows,
we use Equation 7 from \citet{De11} to assign distances to candidate
BHB stars within the colour-colour box outlined in the Left panel of
Figure~\ref{fig:bhb}.

Only one globular cluster has significant non-zero residuals when the
distance modulus $(m-M)_g$ calculated using the BHB distance
calibration described above is compared with $(m-M)_g$ found in the
literature. This is illustrated in the right panel of
Figure~\ref{fig:bhb}. As apparent from the histograms of the $(m-M)_g$
residuals in the right hand side of the panel, the distance modulus
for NGC 2419 needs to be revised $\sim 0.2$ mag upwards, thus making
its heliocentric distance not 84 kpc \citep{Ha96}, but 91
kpc. Recently, \citet{Fe13} presented new calibration of the BHB
absolute magnitude based on theoretical isochrones. Compared to
\citet{De11} these models predict slightly higher ($\sim 0.1$ mag)
luminosity for stars as metal-poor as those in NGC 2419. To test
whether some of the 0.2 mag discrepancy observed here could be
explained away by this effect, we checked the $(m-M)_g$ residuals for
three more star clusters with $[{\rm Fe}/{\rm H}] < -2$, namely M 92,
M15 and NGC 5053. The residuals for the stars in all three clusters
stay firmly on zero. While our slightly higher distance estimate is
mildly inconsistent with the value provided by \citet{Ha96}, we are in
agreement with the most recent RR Lyrae measurement by \citet{Di11}.

\subsection{The Sgr Stream signal in the SDSS data}

\citet{Ma03} and \citet{Ko10} have shown that measuring the properties
of a stellar stream can be highly robust (even with small numbers of
stream members) if the debris mapping is performed in a coordinate
system aligned with the stream. We follow the prescription of
\citet{Ma03} and transform equatorial RA and Dec into a heliocentric
$\tilde{\Lambda}_{\odot}, \tilde{B}_{\odot}$ coordinate system whose
equator is aligned with the Sgr trailing tail. Note, however, that in
this paper $\tilde{\Lambda}_{\odot}$ increases in the direction of Sgr
motion, i.e. opposite to the convention used by
\citet{Ma03}. Likewise, the latitude axis $\tilde{B}_{\odot}$ points
to the Galactic North pole instead of the South pole. This means that
the coordinates given here can be transformed to those in \citet{Ma03}
convention using the following simple equations:
$\Lambda_{\odot}=360^{\circ}-\tilde{\Lambda}_{\odot}$ and
$B_{\odot}=-\tilde{B}_{\odot}$. While this is a minuscule departure
from the previously used notation, it seems more logical to point the
principal axis of the stream coordinate system in the direction of the
stream's motion. \footnote{The Sgr stream coordinate system has its
  North pole in ($\alpha,\delta$)=($303 \fdg 63, 59 \fdg 58$). In
  Appendix~\ref{sec:appendix} we give the equations necessary to
  convert between the equatorial and the Sgr stream coordinate
  systems.}

The top panel of Figure~\ref{fig:data} shows the density of
(primarily) MSTO stars selected using the colour-magnitude cuts
identical to those in \citet{Ko12}. Both ``branches'' of the Sgr
leading tail are visible superposed onto the Virgo overdensity at
$70^{\circ} < \tilde{\Lambda}_{\odot} < 120^{\circ}$. Also visible are the
Orphan stream at $130^{\circ} < \tilde{\Lambda}_{\odot} < 150^{\circ}$ and the
Monoceros ring at $\tilde{\Lambda}_{\odot}\sim 170^{\circ}$. The only
tentative evidence of the presence of the trailing tail in the
Northern SDSS data is the faint overdensity running between the
Branches A and B at $150^{\circ} < \tilde{\Lambda}_{\odot} < 160^{\circ}$ at
constant $\tilde{B}_{\odot} \sim -5^{\circ}$. Note that even though the
contiguous coverage of the main SDSS footprint runs out at
$\tilde{\Lambda}_{\odot} \sim 70^\circ$ and $\tilde{\Lambda}_{\odot} \sim 170^{\circ}$,
within $|\tilde{B}_{\odot}| < 10^{\circ}$ there exist several stripes reaching
as far out as $\tilde{\Lambda}_{\odot} \sim 40^{\circ}$ and $\tilde{\Lambda}_{\odot}
\sim 190^{\circ}$, providing an additional $\sim 50^{\circ}$ along the
stream.

Indeed, the stream is clearly visible at both lower and higher
$\tilde{\Lambda}_{\odot}$, as revealed by the density of the candidate BHB
stars in the middle panel of the Figure. Only the stars that lie
inside the $u-g, g-r$ boundary shown in the Left panel of
Figure~\ref{fig:bhb} (see \citealt{Ya00}) and within $-12^{\circ} <
\tilde{B}_{\odot} < 12^{\circ}$ are included in this picture. The leading tail
can be traced from $\tilde{\Lambda}_{\odot} \sim 35^{\circ}$ to
$\tilde{\Lambda}_{\odot} \sim 120^{\circ}$ with both BHB at $17 < (m-M)_g <
19$ and BS stars at $(m-M)_g > 19.5$. The stream signal traced by the
A-coloured stars dries up much faster than that traced by the MSTO
stars; this is simply the consequence of the declining stream
luminosity and the decreasing distance as explained in \citet{NO10}.

\begin{figure*}
  \centering
  \includegraphics[width=16.5cm]{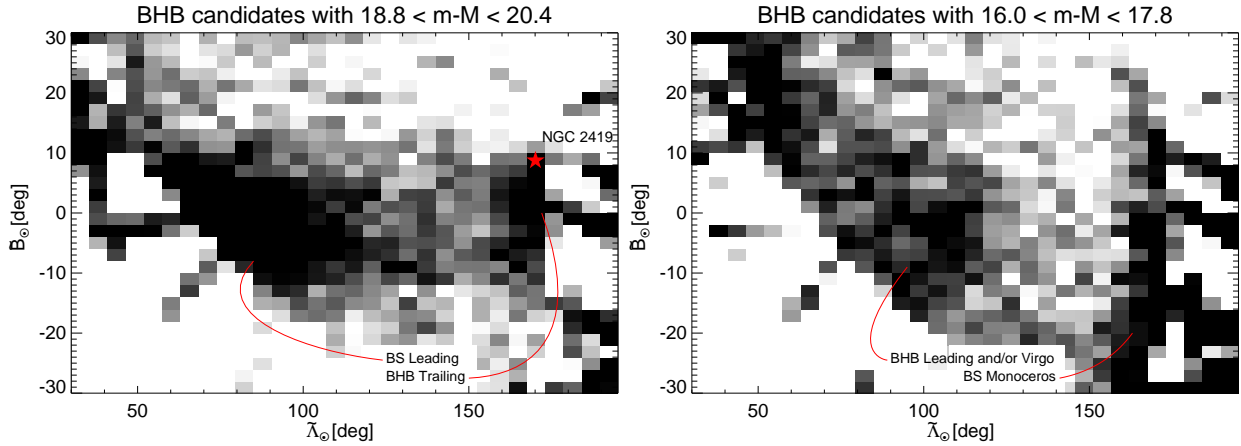}
  \caption[]{\small Sky density of the BHB candidates. \textit{Left:}
    Stars with $18.8 < m-M < 20.4$ (see Middle panel of Figure
    \ref{fig:data}). Both leading tail and trailing tail can be easily
    discerned. Globular cluster NGC 2419 appears projected right at
    the edge of the trailing stream. \textit{Right:} Density of the
    brighter BHB candidate stars. None of the Sgr tails appears
    distinctly visible in this picture, although an over-density is
    detected at $80^{\circ} < \tilde{\Lambda}_{\odot} < 120^{\circ}$, which
    could plausibly be attributed either to the portion of the Sgr
    leading stream falling into this distance range, or to the Virgo
    Stellar Structure. At high $\tilde{\Lambda}_{\odot}$, a narrow vertical
    band parallel to the Galactic disk is visible. The A-colored stars
    in this structure could either be young stars in the
    disk,alternatively they could be the denizens of the Monoceros
    stream.}
   \label{fig:trail_north}
\end{figure*}

The trailing tail is expected to enter the SDSS footprint at high
$\tilde{\Lambda}_{\odot}$. In the middle panel of Figure~\ref{fig:data}, there
are two density enhancements at $\tilde{\Lambda}_{\odot} > 150^{\circ}$: a
broad faint cloud at $16 < (m-M)_g < 17.5$ and a narrow sequence at
$19 < (m-M)_g < 20.5$. To establish which of the two signals is
contributed by the Sgr debris, the 2D histograms of the candidate BHB
stars in the corresponding $(m-M)_g$ ranges are plotted in
Figure~\ref{fig:trail_north}. Dominating the Left panel of
Figure~\ref{fig:trail_north} are two overdensities each ending sharply
at $|\tilde{B}_{\odot}|>10^{\circ}$. The one at lower $\tilde{\Lambda}_{\odot}$
obviously corresponds to the Sgr leading tail, while the short stubby
piece of the stream at higher $\tilde{\Lambda}_{\odot}$ clearly has not
precessed as much as the nearby end of the Branch A (see top panel of
Figure~\ref{fig:data}). The fact that the width of the distribution of
the distant BHB stars matches that of the rest of the Sgr stream has
already been noted by \citep{Ne03}. This picture is distinctly
different from the one seen in the Right panel of the Figure, where
the shape of the distribution of the brighter BHB-like stars bears
little resemblance to the Sgr stream. At around $\tilde{\Lambda}_{\odot} \sim
110^{\circ}$, there appears an over-density that could correspond to
either the brightest of the leading tail BHBs or stars in the Virgo
over-density. Most importantly, at $\tilde{\Lambda}_{\odot} > 160^{\circ}$,
the stars are distributed in a band perpendicular to the direction of
the stream, thus ruling out any connection to Sagittarius.

It is reasonable to conjecture that the debris traced by the distant
BHB stars is contributed by the Sgr trailing tail that passed through
the Galactic disk and is near its apo-centre at $\tilde{\Lambda}_{\odot} \sim
170^{\circ}$. To further test this hypothesis, the line-of-sight (LOS)
velocities $V_{\rm GSR}$ of all SDSS DR8 giant stars in the vicinity
of the equator of the Sgr coordinate system are plotted in the lower
panel of Figure~\ref{fig:data}. The following magnitude, colour,
effective temperature and surface gravity cuts are applied to pick out
the giants:

\begin{equation}
\begin{split}
17 < g < 21.5 \\
0.8 < g-i < 1.7 \\
T_{eff} < 5,500 K\\
1 < \mathrm{log}(g) < 3
\end{split}
\end{equation}

The first (apparent magnitude) cut is highly effective at eliminating
the disk contamination by simply excising the brightest of the stars
from the sample. The rationale for the remaining 3 selection criteria
is illustrated in Figure~\ref{fig:tracers}. As obvious from the
Figure, the effective temperature and the surface gravity cuts simply
isolate the region dominated by the RGB stars. As the Sgr is somewhat
more metal-rich as compared to the rest of the stellar halo, the
additional $g-i$ is used to get rid of the bluer, more metal-poor
RGBs. We argue that while this cut reduces the stream signal somewhat
(as it is known that the Sgr tails contain a modest metal-poor
population), it suppresses the halo contamination even more.

\begin{figure}
  \centering
  \includegraphics[width=0.98\linewidth]{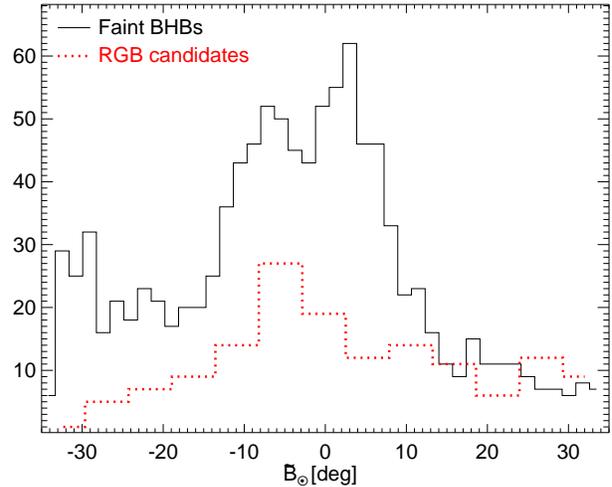}
  \caption[]{\small Across-stream profiles of the candidate members of
    the Sgr trailing stream in the North. The faint BHB stars (solid)
    and the spectroscopic RGBs (red dotted) in the range $145^{\circ}
    < \tilde{\Lambda}_{\odot} < 170^{\circ}$ are shown. The BHB
    distribution appears double-peaked, while the more metal-rich RGB
    profile peaks at $\tilde{B}_{\odot} \sim -5^{\circ}$. This
    behaviour is reminiscent of the recent measurement of the trailing
    debris in the Galactic South by \citet{Ko12}, who detect the
    similar metallicity difference between the stream components at
    positive and negative $\tilde{B}_{\odot}$.}
   \label{fig:profiles}
\end{figure}

Within $-9^{\circ} < \tilde{B}_{\odot} < 9^{\circ}$, the distribution
of the stars remaining after the above cuts is dominated by the two
narrow sequences. Starting at $V_{\rm GSR} \sim 50~{\rm km/s},
\tilde{\Lambda}_{\odot} \sim 60^{\circ}$ is the Sgr leading tail, with
its velocity gently decreasing with longitude down to $V_{\rm GSR}
\sim -120$ km/s at $\tilde{\Lambda}_{\odot} \sim 130^{\circ}$, where
it seems to flatten. Wherever the datasets overlap, the leading tail
kinematics reported here are in good agreement with the velocities of
the M giants reported by \citet{La04, La05}. The other similarly
narrow overdensity runs from $\tilde{\Lambda}_{\odot} \sim
300^{\circ}$ to $\tilde{\Lambda}_{\odot} \sim 210^{\circ}$ where it is
interrupted briefly by the Galactic disk. It then appears to re-emerge
at $\tilde{\Lambda}_{\odot} \sim 185^{\circ}$ and $-50 < V_{\rm GSR} <
-100$ km/s, from where it rises to $V_{\rm GSR} \sim 100$ km/s at
$\tilde{\Lambda}_{\odot} \sim 140^{\circ}$. There is no doubt that the
Southern portion of this signal is the Sgr trailing debris as
previously seen using the M giants by \citet{La04, La05}, as well as
various other tracers in the SDSS \citep[e.g.][]{Ya09, Ko12}. The
Northern extension, below $\tilde{\Lambda}_{\odot} \sim 190^{\circ}$
is new. There are several clues that the Northern and the Southern
velocity signals are perhaps related. First, notwithstanding the small
gap due to the disk, the radial velocity data in the South and the
North seem to link smoothly. Additionally, the line-of-sight velocity
dispersions of the two pieces look rather similar.  More importantly,
Figure~\ref{fig:profiles} confirms that the RGB velocity signal at
$140^{\circ} < \tilde{\Lambda}_{\odot} < 190^{\circ} $ is also limited
to the Sgr plane, and hence, unlikely to be due to a different stellar
halo sub-structure that enters the selection box $-9^{\circ} <
\tilde{B}_{\odot} <9^{\circ}$ spuriously. The Figure shows the
across-stream profile of the faint BHBs with $18.8 < m-M < 20.4$ (see
also Left panel of Figure~\ref{fig:trail_north}), as well as the
profile of the RGB stars giving rise to the velocity overdensity in
question as seen in the Bottom panel of Figure~\ref{fig:data}. Both
the BHB and the RGB stars selected to produce these 1D distributions
are restricted to lie in the range $145^{\circ} <
\tilde{\Lambda}_{\odot} < 170^{\circ}$. This is because at higher
$\tilde{\Lambda}_{\odot}$, the SDSS footprint is incomplete, and at
lower $\tilde{\Lambda}_{\odot}$, the stream signal quickly runs
out. The BHB profile appears to have two peaks indicating that the
stream possibly has two distinct but overlapping components of similar
width. The distribution of the spectroscopic RGB candidates peaks at
$B\sim -5^{\circ}$. This behaviour is in excellent agreement with the
findings of \citet{Ko12} who show that in the Southern hemisphere, the
Sgr trailing tail is bifurcated, and the stream component at lower $B$
contains a more metal-rich population, while the component at higher
$B$ is dominated by the old and metal-poor stars.

To summarize, each of the three panels of Figure~\ref{fig:data} shows
the properties of a different stellar tracer inside the Sgr
stream. The underlying logic that links the distribution of MSTO stars
on the sky with the distance evolution of the BHBs and the kinematics
of the RGB stars is threefold. First, the density enhancements in all
three stellar populations are limited to the Sgr orbital plane as
represented by the rotated coordinate system similar to that of
\citet{Ma03}. Second, both the BHB and the RGB overdensities are
narrow enough to match the cross-section of the stream as measured on
the celestial sphere. Third, as a function of the longitude
$\tilde{\Lambda}_{\odot}$, the RGB kinematics evolves in agreement with the
distance gradients as traced by the BHB stars. In particular, the LOS
velocity of both the leading and the trailing tail goes through zero
in the vicinity of the respective apo-centre.

\subsection{Trailing arm: Connecting the dots}

\begin{figure*}
  \centering
  \includegraphics[width=17cm]{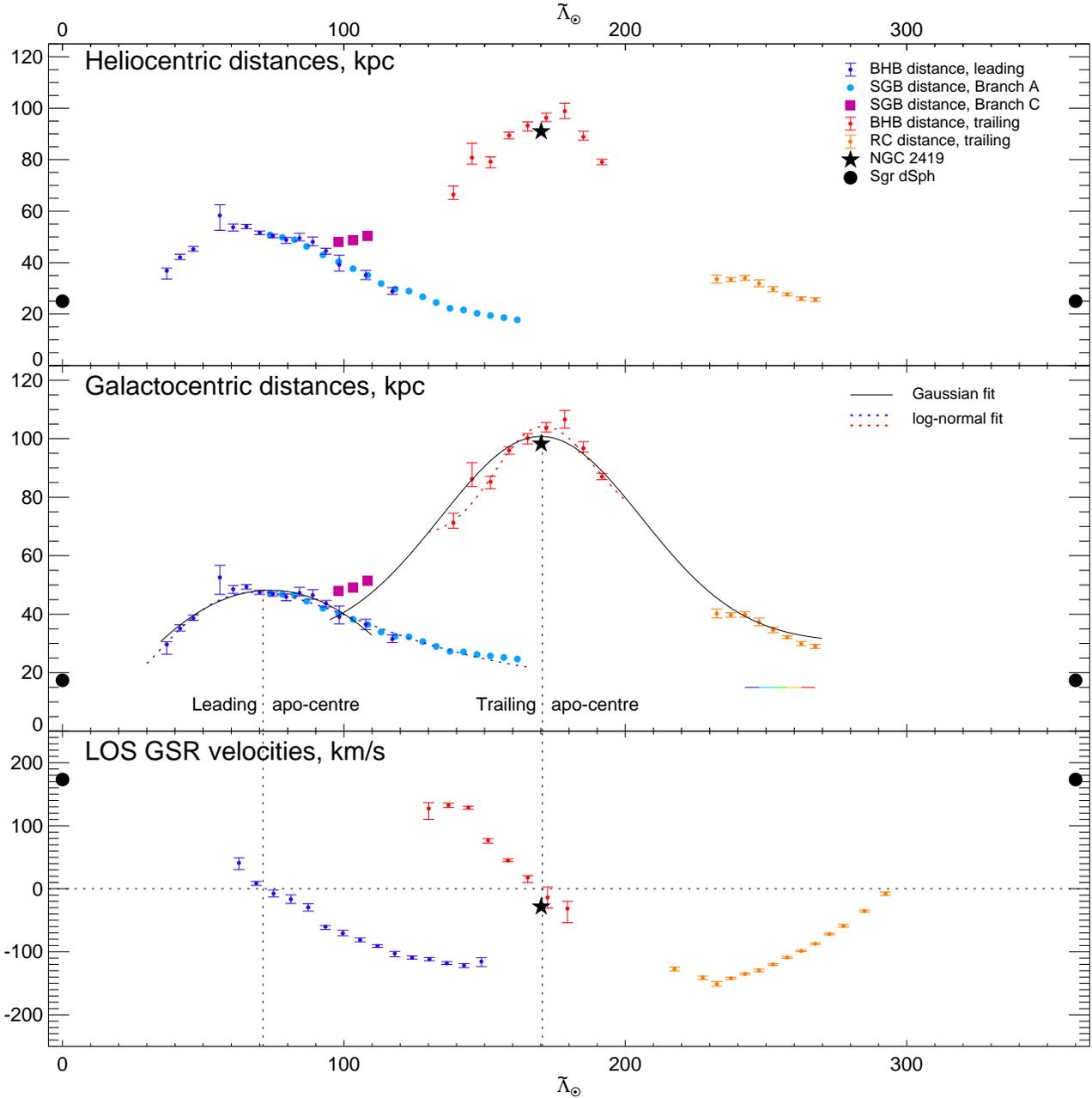}
  \caption[]{\small Distance and velocity measurements of the Sgr
    stream.\textit{Top:} Violet (red) data-points with error bars show
    the centroid of the heliocentric distance of the stream debris at
    given longitude $\tilde{\Lambda}_{\odot}$ for the leading
    (trailing) tail. Blue (magenta) filled circles (squares) are
    SGB-based Branch A (C) distance measurements from \citet{Be06}
    increased by 0.15 mag to match the BHB signal. Orange data-points
    with error-bars are RGB-based distance measurements from
    \citet{Ko12} increased by 0.35 mag to correct for the reddening
    towards the progenitor. A black star marks the location of the
    globular cluster NGC 2419. \textit{Middle:} Galactocentric stream
    distances. The stream is assumed to be at $B=0^{\circ}$
    everywhere. While this will bias the run of distances for the
    individual branches of the leading arm at
    $\tilde{\Lambda}_{\odot}>150^{\circ}$, this is a very reasonable
    approximation for the debris around both apo-centres. Violet and
    red solid curves show the log-normal fits to the data, while black
    solid curves represent pure Gaussian models. Dotted lines mark the
    location of the leading and trailing apo-centres. \textit{Bottom:}
    Measurements of the line-of-sight velocity $V_{\rm GSR}$ along the
    stream. The velocity centroids are those based on the SDSS giants
    stars as presented in Tables~\ref{tab:lead_vel},
    ~\ref{tab:trail_vel} and ~\ref{tab:trail_vel_south}. Note that the
    stream velocity appears to go through zero in the vicinity of the
    apo-centre.}
   \label{fig:data_measure}
\end{figure*}

Given the clarity of the stream signal in both distance and the
velocity domain, it is straightforward to measure the change in the
centroid of the Sgr debris as a function of $\tilde{\Lambda}_{\odot}$. To this
end, we build the model of the Galactic foreground and the Sgr stream
as follows. At each $\tilde{\Lambda}_{\odot}$, the latitudinal probability
that a star belongs to the stream is Gaussian, while the probability
of belonging to the foreground changes linearly with latitude
$\tilde{B}_{\odot}$. Therefore, there are 4 unknowns describing the centre and
the width of the Gaussian and the slope and the normalization of
foreground. The sets of 4 model parameters are allowed to evolve
freely from bin to bin in longitude $\tilde{\Lambda}_{\odot}$. In each bin,
the maximum likelihood model is sought using the individual BHB
distance modulus values and the RGB velocity values with a brute-force
grid search implemented in IDL. As is obvious from the
Figure~\ref{fig:data}, the foreground density field is not exactly
linear as a function of $\tilde{B}_{\odot}$. Therefore, when fitting the
distance modulus signal, only values within $\pm 1$ mag of the
tentative stream centroid were used. Similarly, while modelling the
velocity, only the values inside $\pm 150$ km/s margin were used. The
width of the bin in $\tilde{\Lambda}_{\odot}$ was chosen through trial and
error to maintain reasonably high longitudinal resolution while having
enough signal-to-noise per bin. The distances are measured in bins
that are $4 \fdg 7$ ($6 \fdg 6$) wide, for the leading (trailing) arm
respectively. The velocities are measured in bins that are $6 \fdg 2$
($7 \fdg 05$) wide, for the leading (trailing) arm respectively. The
distance and velocity centroids with the associated uncertainties are
reported in Tables~\ref{tab:lead_dist},~\ref{tab:trail_dist} and
\ref{tab:lead_vel},~\ref{tab:trail_vel} and ~\ref{tab:trail_vel_south}
respectively and shown in Figure~\ref{fig:data_measure}.

Top and middle panels of Figure~\ref{fig:data_measure} show the
distances to the centroids of the BHB stars in the Sgr stream measured
as described above. Also shown are the distances measured using the
stream's Sub Giant Branch (SGB) stars, as reported by
\citet{Be06}. Note, however, that there is an important discrepancy
between the SGB and the BHB measurements. The SGB distance moduli
$(m-M)_{\rm SGB}$ needs to be increased by 0.15 mag to match the
values of BHB distances at the same locations. The SGB measurements in
\citet{Be06} are differential with respect to the main body of Sgr
dSph and can be placed on the absolute scale only by assuming the
overall metallicity of the Stream. \citet{Be06} suggested that the
Stream's stellar population resembles closely that of the remnant and
hence assigned the same $i$-band magnitude to the SGB in the Stream
and the Sgr core. The offset between the SGB and BHB stars measured
here implies that the SGB of the Stream is 0.15 magnitudes brighter
than the SGB of the remnant and hence more metal poor. This is in good
agreement with the recent measurements of metallicity gradients along
the Sgr stream (e.g. \citealt{Bela06, Ch07,
  NO10}). \footnote{\citet{NO10} report a reasonable match between the
  SGB and the BHB distances along the Sgr stream, however they assume
  fixed absolute magnitude for BHB stars of $M_g=0.7$, while most of
  the redder BHB stars have $M_g\sim0.5$.} Alternatively, simply
assuming the distance to the Sgr remnant of 26 kpc instead of 24 kpc
produces a similar offset in $(m-M)$.

\begin{figure*}
  \centering
  \includegraphics[width=0.98\linewidth]{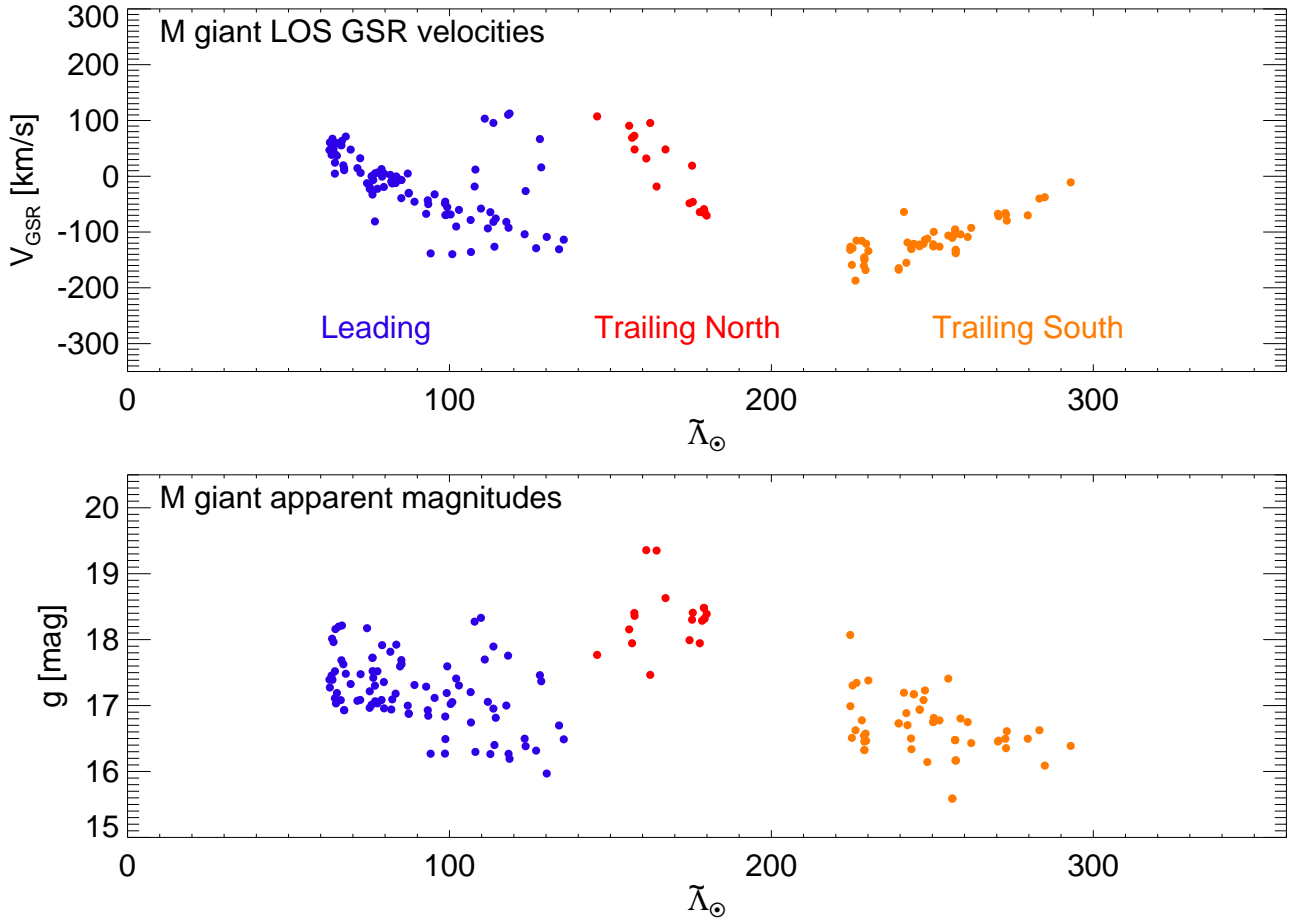}
  \caption[]{\small \textit{Top:} Radial velocities of the M giant
    candidate stars with $-9^{\circ} < \tilde{B}_{\odot} < 9^{\circ}$
    selected using the criteria specified in equation~2 as a function
    of $\tilde{\Lambda}{\odot}$. The color-coding is used to emphasize
    the obvious velocity groups. The radial velocity trends match well
    those presented in Figures~\ref{fig:data} and
    ~\ref{fig:data_measure}. \textit{Bottom:} Apparent $g$ band
    magnitude of the selected M giant candidates as a function of
    $\tilde{\Lambda}_{\odot}$. The distance trends are similar to
    those reported in Figure~\ref{fig:data_measure}. The large spread
    is the result the strong dependence of the M giant absolute
    magnitude on the temperature (color), the metallicity and the
    age. See text for the estimates of the M giant distances.}
   \label{fig:mgiants1}
\end{figure*}

With the accurate distance measurements of the trailing debris
covering $140^{\circ} < \tilde{\Lambda}_{\odot} < 190^{\circ}$ reported here,
it is easy to ``connect the dots'' and conclude that the most natural
explanation for the so-called Branch C \citep[see e.g.][]{Be06,Fe06},
confirmed most recently by \citet{Co10} with Red Clump giants, is
simply the extension of the trailing tail into the range of lower
$\tilde{\Lambda}_{\odot}$ where it overlaps with the leading tail at
$\tilde{\Lambda}_{\odot} \sim 100^{\circ}$. The trailing stream lies further
out in the halo and follows a steeper distance gradient than predicted
by any Sgr disruption model to date. For example, in the model of
\citet{LM10} the crossing of the leading and trailing tails in the
North happens at RA$\sim 150^\circ$ or $\tilde{\Lambda}_{\odot} \sim
140^\circ$, thus missing it by some $40^\circ$. The available
kinematics of the trailing tail lend support to our
interpretation. The lower panel of Figure~\ref{fig:data_measure} shows
that as the debris reach the apo-centre at $\tilde{\Lambda}_{\odot} \sim
170^{\circ}$, the LOS velocity changes sign and goes through zero. The
maximum LOS velocity is reached at a point where the line of sight is
best aligned with the stream. In the trailing arm velocity data, there
is a clear indication of the plateau at $\tilde{\Lambda}_{\odot} \sim
135^{\circ}$. The LOS velocity is then expected to drop to zero close
to the peri-centre, which for the trailing tail seems to lie not too
far from the point of the crossing with the leading tail.

If the above interpretation is correct, then the Sgr trailing debris
are flung out as far as 100 kpc away from the Galactic centre. This in
turn implies a difference of $\sim 50$ kpc between the leading and the
trailing apo-centres, which is not predicted by any of the current Sgr
disruption models. While the orbital precession is sensitive to the
global properties of the potential probed by the orbit, the difference
in apo-central distances is also the consequence of the offset in
energy and angular momentum of the debris at the moment of stripping,
which for systems like Sgr, happens predominantly at peri-centre. A
larger offset can be either a result of a steeper potential or a
larger tidal diameter of the satellite, or indeed, a combination of
both.

\subsection{The M giant acid test}

M giants are scant at high Galactic latitudes. Compared to the bulk of
the Milky Way stellar halo, these stars are typically much younger
(somewhere in the range of 4 to 8 Gyr) and more metal-rich.  Given
their distinct infrared and ultraviolet colors, M giants can be
identified easily, and provided the redder and therefore more
metal-rich sub-sample is selected, will generally suffer very low
levels of contamination. As exemplified by the studies of \citet{Ma03}
and \citet{Ya09}, M giants are ideal markers to track down the Sgr
tidal debris across the sky. Crucially, for our study, compared to the
older and the more metal-poor BHBs and the RGBs, they represent a
well-studied, largely complementary stellar tracer population with
high purity.

As Figure~\ref{fig:tracers} demonstrates, in the SDSS spectroscopic
dataset, M giant candidates can be picked out using the following
selection criteria.

\begin{equation}
\begin{split}
4,000 < T_{eff} < 4,600 K\\
1 < \mathrm{log}(g) < 2.2\\
2.7 < u-g < 3.5\\
-2 < [Fe/H] <0
\end{split}
\end{equation}

Figure~\ref{fig:mgiants1} shows two distributions of the M giant
candidate stars with $-9^{\circ} < \tilde{B}_{\odot} < 9^{\circ}$ selected
using the above criteria. The data-points are split into three groups
according to the range of $\tilde{\Lambda}_{\odot}$ they inhabit, as indicated
by the color coding. Top panel of Figure~\ref{fig:mgiants1} gives the
M giant locations on the plane of the line-of-sight velocity
(corrected for the solar reflex motion) and the stream longitude
$\tilde{\Lambda}_{\odot}$. It is evident that the absolute majority of the
stars plotted do belong to the Sgr tidal tails. Indeed the
contamination is minimal as there are only few M giants in the range
$100^{\circ} < \tilde{\Lambda}_{\odot} < 140^{\circ}$ that deviate from the
radial velocity trends presented in Figure~\ref{fig:data_measure}. In
fact, perhaps some of these ``contaminating'' stars are part of the
Branch C of the stream, i.e. the continuation of the trailing tail to
lower $\tilde{\Lambda}_{\odot}$. Most importantly, the M giant radial velocity
signature at $140^{\circ} < \tilde{\Lambda}_{\odot} < 190^{\circ}$ matches
perfectly the RGB kinematics presented in the lower panel of
Figure~\ref{fig:data_measure}.  For example, the M giant radial
velocity signal goes through zero around $\tilde{\Lambda}_{\odot} \sim
170^{\circ}$.

\begin{figure}
  \centering
  \includegraphics[width=0.95\linewidth]{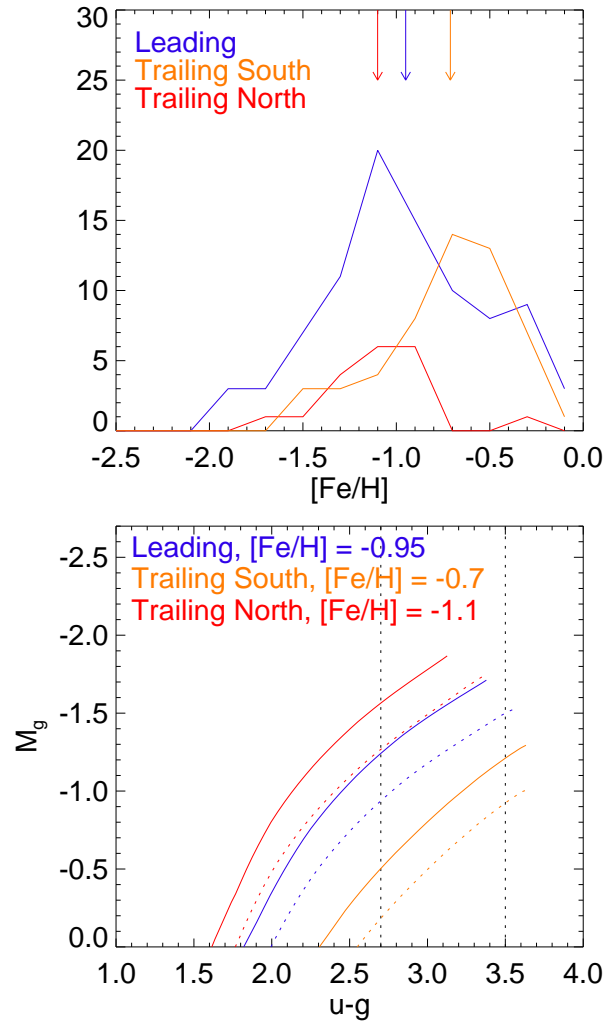}
  \caption[]{\small \textit{Top:} Metallicity distributions of the M
    giant stars in the three $\tilde{\Lambda}_{\odot}$ ranges, as
    indicated in Figure~\ref{fig:mgiants1}. The arrows give the mean
    metallicity values for each M giant sub-sample. Note, that there
    is a significant downward shift in mean $[Fe/H]$ between the
    Southern (orange) and the Northern (red) detections of the
    trailing tail. The $[Fe/H]$ distribution of the leading tail
    (blue) agrees better with the distant trailing debris (red); yet
    there is a 0.15 dex difference in mean metallicities of the two M
    giant groups.  \textit{Bottom:} Dartmouth theoretical isochrones
    \citep{Do08} shown for the three mean M giant metallicities for 4
    Gyr (solid) and 8 Gyr (dotted) populations. Note that even 0.15
    dex difference in $[Fe/H]$ can result in $\sim$0.5 mag difference
    in absolute magnitude.}
   \label{fig:mgiants2}
\end{figure}

An idea of the heliocentric distance of the stream's M giants can be
gleaned from the Bottom panel of Figure~\ref{fig:mgiants1}. Here, the
apparent $g$ band magnitude of the selected stars is plotted as a
function of the longitude $\tilde{\Lambda}_{\odot}$. While the
data-points follow the general distance trends shown in the Top panel
of Figure~\ref{fig:data_measure}, the scatter is substantial. This is
because the intrinsic luminosity of an M giant star varies
significantly with its temperature (and hence color), as well as the
metallicity and the age. Using the BHB stars in the Sgr stream around
$\tilde{\Lambda}_{\odot} \sim 75^{\circ}$, \citet{Ya09} obtained $M_g
= -1$ for the stream's M giants. Using this simple absolute magnitude
calibration we can obtain a crude estimate of the helio-centric
distance to the 19 M giant candidate stars in the range $140^{\circ} <
\tilde{\Lambda}_{\odot} < 190^{\circ}$. Their mean $g$ band apparent
magnitude is 18.4 and the dispersion is 0.4 mag. If there was no
strong metallicity/age gradient along the stream, applying the
absolute magnitude calibration of \citet{Ya09} leads to the distance
estimate of 75$^{+15}_{-10}$ kpc. This is slightly lower but overall
consistent with the distance estimate to the BHB stars in the same
range of $\tilde{\Lambda}_{\odot}$. However, the assumption of the
zero gradient in the stream's stellar populations is perhaps too naive
as shown in Figure~\ref{fig:mgiants2}. The top panel of the Figure
shows the distributions of the metallicity (as derived by the SDSS
SSPP pipeline) of the M giant stars selected to lie in the three
$\tilde{\Lambda}_{\odot}$ ranges as indicated in
Figure~\ref{fig:mgiants1}. There is a pronounced gradient in $[Fe/H]$
between the Southern and the Northern trailing debris, with the mean
metallicity shifting from -0.7 to -1.1. The metallicity distributions
of the leading tail and the distant trailing tail agree much better,
but the average values of $[Fe/H]$ still differ by 0.15 dex. As
illustrated in the lower panel of Figure~\ref{fig:mgiants2}, even 0.15
dex lower metallicity can result in $\sim$ 0.5 mag larger distance
modulus. This would imply that the average distance to the faint M
giant stars could be closer to $\sim 95$ kpc.

Overall, the M giants provide strong support to the picture drawn with
other SDSS stellar tracers. In particular, in the portion of the sky
in question, i.e. in the range of $140^{\circ} < \tilde{\Lambda}_{\odot} <
190^{\circ}$, the M giants, BHBs and RGB follow essentially identical
velocity and distance trends.

\section{NGC 2419}
\label{sec:2419}

\begin{figure*}[t]
  \centering
  \includegraphics[width=17cm]{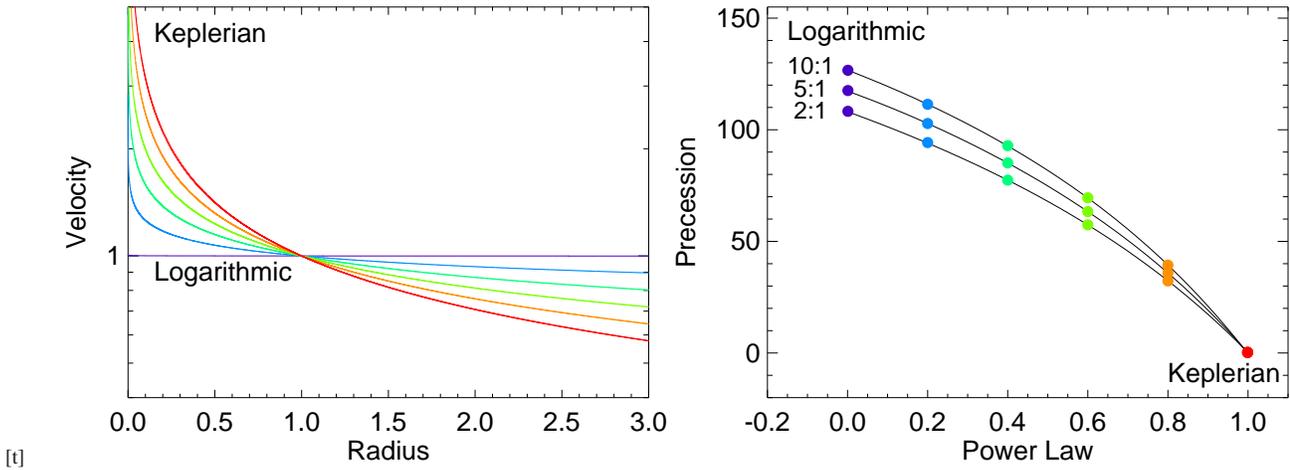}
  \caption[]{\small \textit{Left:} Circular velocity curves in
    power-law potentials as a function of radius. \textit{Right:}
    Apo-centre to apo-centre precession in power-law potentials as a
    function of the power for three different eccentricities. For each
    potential (colour-coded according to power-law index), the
    precession angle in degrees is shown for three orbits with the
    ratio of apo-centric to peri-centric distance of 10:1, 5:1 and
    2:1.}
   \label{fig:precession_powerlaw}
\end{figure*}

NGC 2419 is rather unusual for a Milky Way globular cluster. Together
with $\omega$Cen, it deviates from the size-luminosity relation obeyed
by other Galactic globulars, as noted, for example, by
\citet{Va04}. NGC 2419 appears to have a relaxation time well in
excess of Hubble time. In other words, it is much too extended for its
stellar mass, which by itself makes it one the most luminous GCs in
the Galaxy. \citet{Va04} and others have noted the resemblance between
NGC 2419, $\omega$ Centauri and M54 and suggested that these clusters
could have formed and evolved in dwarf satellites of the Milky Way
which were then destroyed during accretion leaving the fleshy clusters
intact in the gravitational field of the Milky Way. Accordingly, the
search is on for evidence of the past existence of the globular's
parent galaxy. \citet{Va04} predict a spread, albeit possibly quite
small, in the metallicities of the cluster's member stars. Most
recently, \citet{Coh10} seem, at first glance, to have detected
exactly that: a small, but measurable spread in the CaT equivalent
widths of several tens of the RGB stars in NGC 2419, which could be
interpreted as the spread in $[{\rm Fe}/{\rm H}]$ values around the
mean of $-2.1$. However, in the follow-up high resolution study,
\citet{Co11} and \citet{Co12} find no detectable spread in $[{\rm
    Fe}/{\rm H}]$, but unusually high depletion of Mg and a bizarre
anti-correlation between Mg and K. They conclude by stating that no
nucleosynthetic source is capable of explaining the chemical makeup of
NGC 2419.

\begin{figure*}
  \centering
  \includegraphics[width=17cm]{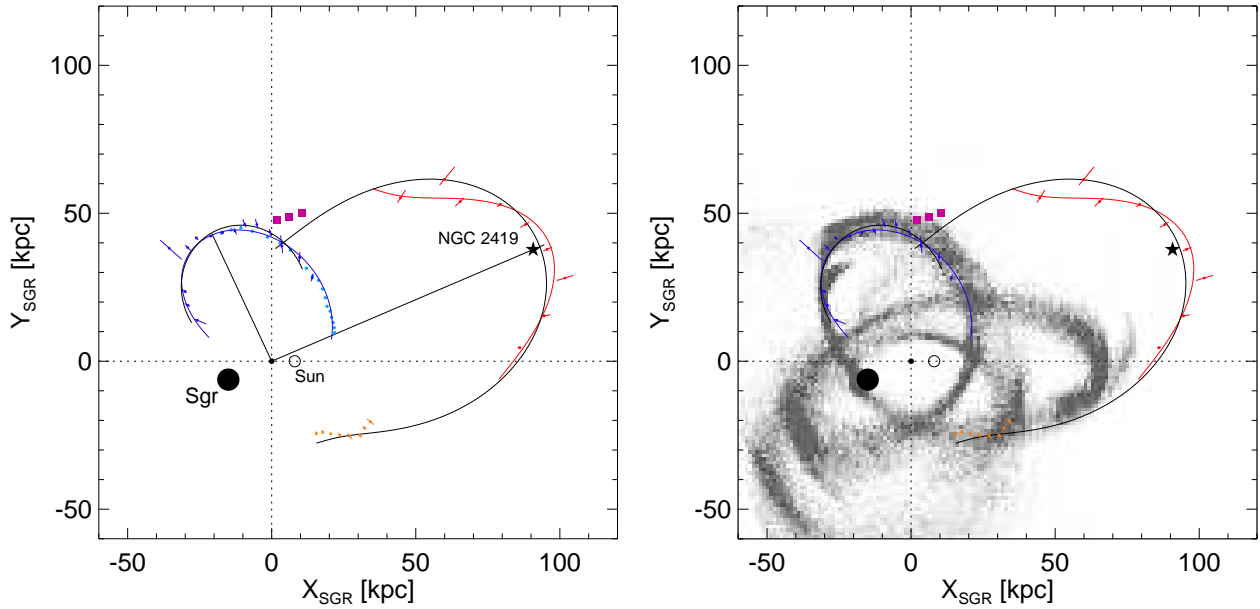}
  \caption[]{\small Stream precession in the plane of the Sgr
    orbit. The plane chosen has its pole at Galactocentric $l_{\rm
      GC}=275^{\circ}$ and $b_{\rm GC}=-14^{\circ}$. All symbols,
    colours and curves are identical to those in Figure
    \ref{fig:data_measure}. \textit{Left:} Note that the actual
    Galactocentric orbital precession of $93^{\circ}$ is slightly
    lower than the difference between the heliocentric apo-centre
    phases from Figure \ref{fig:data_measure}. \textit{Right:}
    Comparison with the Sgr disruption model by \citet{LM10} shown as
    grey-scale density. Note that in the logarithmic halo used in the
    model, the orbital precession is $\sim 120^{\circ}$ and the
    trailing apo-centre lies in the Galactic disk at $\tilde{B}_{\rm GC}^{\rm
      T} \sim 0^{\circ}$ and distance of $R^{\rm T} \sim 65$ kpc. This
    should be contrasted with the new measurement of $\tilde{B}_{\rm GC}^{\rm
      T} \sim 23^{\circ}$ and $R^{\rm T}=102.5 \pm 2.5$ kpc.}
   \label{fig:data_inplane}
\end{figure*}

Of course, any sign of the tidal debris that can be traced back to the
vicinity of NGC 2419 would be a giveaway just as well. \citet{Ne03}
show that, within the Sgr debris plane, an overdensity of BHBs at
distances similar to that of NGC 2419 can be seen in the SDSS
photometric data. The significance of this BHB overdensity was
confirmed recently by \citet{Ru11}. Given the proximity of the stream
and the cluster, \citet{Ne03} speculate that the cluster was once part
of the Sgr galaxy. \citet{Ca09} on the other hand link the cluster
with the Virgo Stellar Stream (which is, most likely, a part of the
Virgo Over-Density) on the basis of the cluster's proximity to the
very eccentric orbit predicted from the measured mean radial velocity
and the proper motion of the stream. In this paper, we have shown that
NGC 2419 does lie close to the plane of the Sgr tidal
debris. According to Figure~\ref{fig:trail_north}, NGC 2419 is
situated near the edge of the stream at $\tilde{B}_{\odot} \sim
9^{\circ}$. Figures~\ref{fig:data_measure} and~\ref{fig:data_inplane}
show that NGC 2419 is located right at the apo-centre of the Sgr
trailing debris, more than 100 kpc away from the remnant. Finally, the
cluster's LOS velocity matches that of the receding trailing debris
reaching its apo-centre. If this globular was indeed part of the Sgr
dwarf in the distant past, it might not be too surprising that
presently it is found further away from both the centroid of the
stream and the remnant itself. Had it lived closer to the central
parts of Sgr, chances are it would have fallen into the centre of the
dwarf galaxy due to the dynamical friction.

\begin{figure*}
  \centering
  \includegraphics[width=0.98\linewidth]{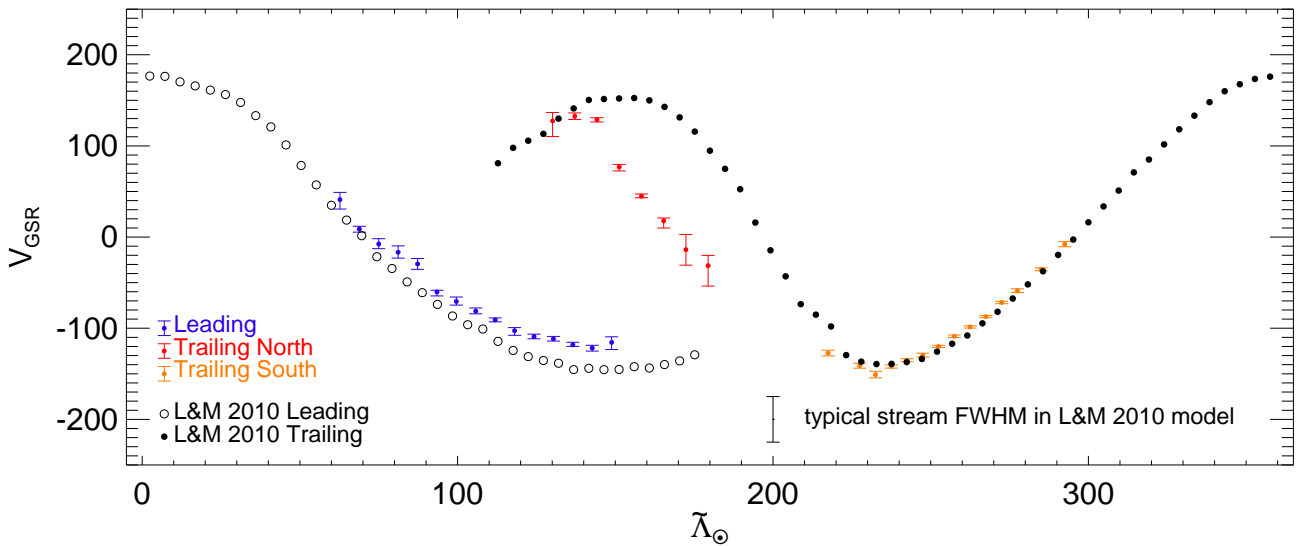}
  \caption[]{\small Measurements of the Sgr stream radial velocity
    (colored data points) and the predictions of the \citet{LM10}
    disruption model (filled and empty circles). In each bin of
    $\tilde{\Lambda}_{\odot}$, the centroid of the model debris is
    given by the median particle velocity. The agreement between the
    data and the model is perfect in the South. The kinematics of the
    leading tail in the North is also reproduced fairly well, although
    there seems to be a systematic offset of order of $\sim$20 km
    s$^{-1}$. This offset is however smaller than the breadth of the
    model tidal debris as indicated by the stand-alone black
    error-bar. The most significant discrepancy is between the distant
    trailing tail data (red) and the model prediction (filled circles)
    in the range $130^{\circ} < \tilde{\Lambda}_{\odot} <
    190^{\circ}$.}
   \label{fig:compare_vel}
\end{figure*}

There is a slight hint of bimodality in the distribution of the
distant BHBs at around $\tilde{\Lambda}\sim 160^{\circ}$ in the Left panel of
Figure~\ref{fig:trail_north}. This is confirmed in the across-stream
distribution of the BHBs shown in Figure~\ref{fig:profiles}. The width
of this part of the trailing stream on the sky is $\sim
20^{\circ}$. This is consistent with the detections of the Sgr debris
everywhere else on the celestial sphere. However, everywhere else,
these $\sim 20^{\circ}$ are made up of two distinct components: Branch
A and B around the North Galactic Cap \citep{Be06}, bright and faint
trailing stream in the South \citep{Ko12}. As judged by the latitude
of NGC 2419 in the Sgr coordinate system, it lies much closer to the
debris plane defined by the faint companion to the trailing
tail. According to \citet{Ko12} this newly-found stream is metal-poor
which helps to strengthen its link to NGC 2419 which possesses $[{\rm
    Fe}/{\rm H}] \sim -2.1$.

There are examples of globular clusters as extended as NGC 2419
outside the Milky Way. \citet{Brodie11} discovered a population of
low-surface brightness counterparts to Ultra Compact Dwarfs in M87,
spanning a similar range in size and luminosity. \citet{Hu05}
discovered a population of faint ($M_V\gtrsim -7$) and fuzzy globulars
with typical sizes $\sim30$pc in the neighboring spiral
M31. Interestingly, \citet{Ma10} showed that all but one of these
extended clusters in Andromeda's outer halo lie - in projection -
within the known tidal streams. This points to the dwarf satellites of
luminous galaxies as the birth place of extended clusters, at least
the ones similar to those discovered by \citet{Hu05}. Some evidence to
support this hypothesis has been published recently. \citet{Da09}
presented a discovery of an extended globular cluster in a low
luminosity dwarf ellipitical member of the Virgo group. There are also
hints of bimodality in the size distribution of globular clusters in
dwarf galaxies in the Local Volume \citep{Ge09}.

With its unusually high (for a Milky Way globular cluster) luminosity
of $M_{\rm V}=-9.4$, NGC 2419 may have more in common with so-called
Ultra-Compact Dwarf (UCD) galaxies \citep{Dr00} than with extended
globular clusters. However, not only are the UCDs at least a magnitude
more luminous, they also appear to have somewhat inflated velocity
dispersions, as evident from the inferred mass-to-light ratios of
$\sim 5-10$ similar to that of the nuclei of dwarf ellipticals in
clusters of galaxies (e.g. \citealt{Ge02}). It is therefore not
surprising that one of the more popular UCD formation scenarios
involves tidally stripping a dwarf elliptical. As of today, the
progenitors of the UCDs have not yet been established with
certainty. Also, as \citet{Ha05} show, it is close to impossible to
assign an astrophysical class to an object in a certain range of
luminosity, size and velocity dispersion as quite a few
``dwarf-globular transition objects'' exist with properties in between
globular clusters and dwarf galaxies. Even though a whole zoo of such
transition objects have been uncovered elsewhere, in the Milky Way NGC
2419 is in a class of its own: too luminous but not dense enough for a
globular, close to the faint end of UCDs, yet, as has been shown by
\citet{Ba09}, it can not boast a mass-to-light ratio out of ordinary.

\begin{figure*}
  \centering
  \includegraphics[width=17cm]{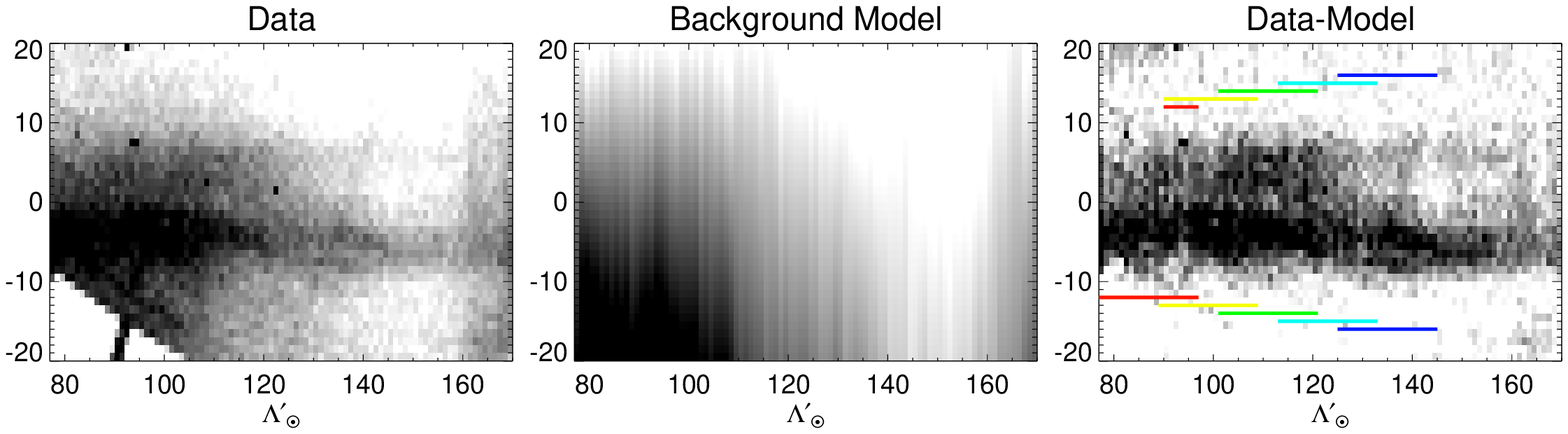}
  \caption[]{\small Foreground model subtraction for the debris plane
    analysis.\textit{Left:} Density of the MSTO stars in the
    coordinate system with the pole at $l=99 \fdg 7$, $b=13\fdg7$ to
    ensure that both Branch A and B lie at approximately constant
    latitude. \textit{Middle:} Foreground stellar density modelled as
    constant slope for each column of the density presented in Left
    panel. The slope is allowed to vary as a function of the longitude
    $\Lambda_{\odot}^{\prime}$ \textit{Right:} Branches A and B
    revealed after the foreground subtraction. Coloured lines show the
    sections of the stream used for the debris plane calculation (see
    Figure \ref{fig:pole}).}
   \label{fig:data_pole}
\end{figure*}

\begin{figure*}
  \centering
  \includegraphics[width=17cm]{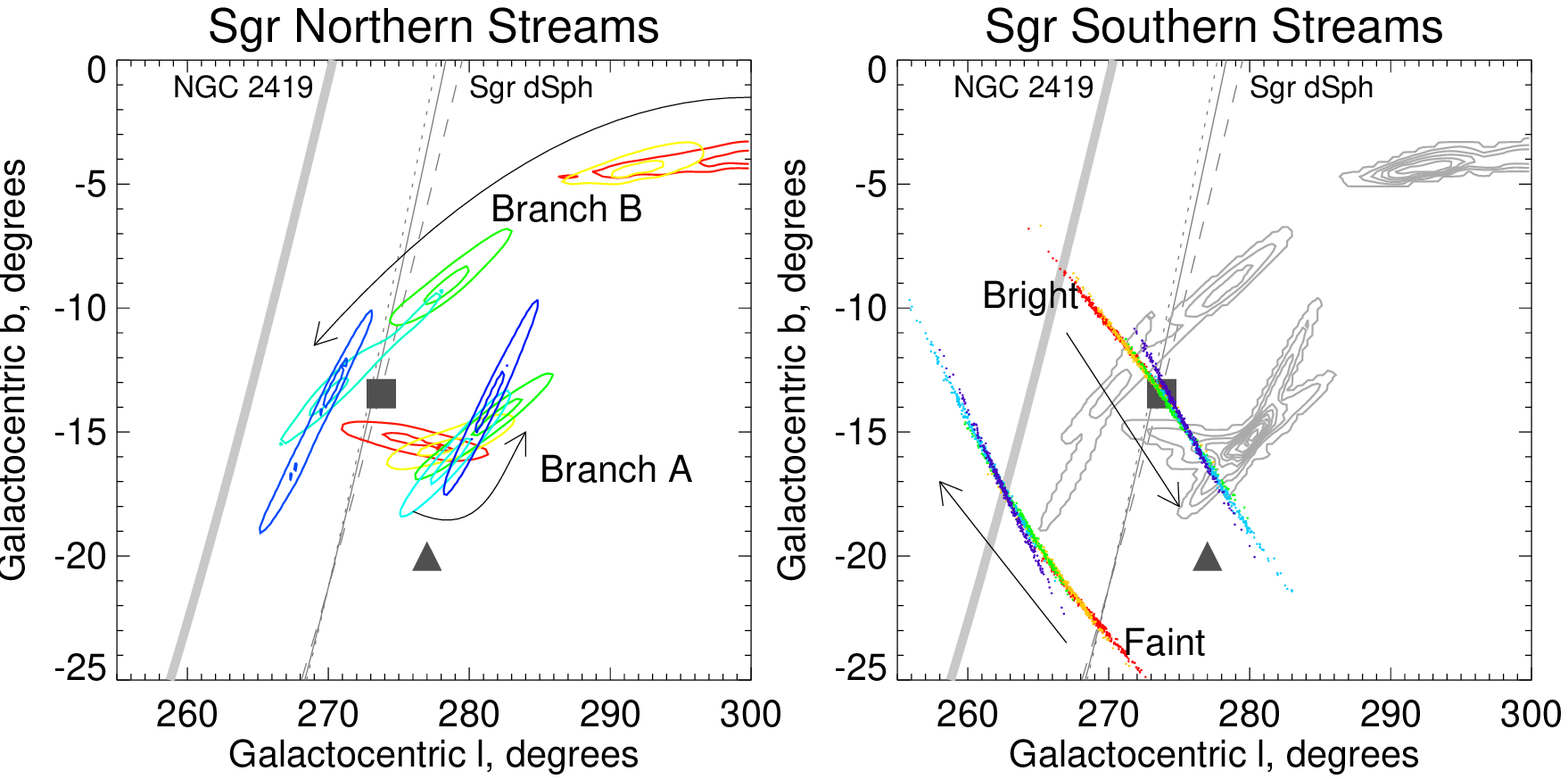}
  \caption[]{\small Precession of the plane of the Sgr debris in
    Galactocentric $l$ and $b$ coordinates. Three black lines crossing
    each panel show the great circles drawn by the poles of the planes
    passing through the current position of the Sgr remnant. Solid is
    for the Sgr dSph heliocentric distance of 25 kpc, while dotted
    (dashed) is for the distance increased (reduced) by 5 kpc. Thick
    light grey line gives the great circle of planes passing through
    NGC 2419. \textit{Left:} Two sets of contours mark the peaks in
    the pole density (99.4\% and 99.9\%) for the five over-lapping
    sections of the two branches of the leading stream (as indicated
    in the Right panel of Figure \ref{fig:data_pole}). Filled square
    (triangle) shows the debris plane of the trailing (leading) stream
    as reported by \citet{Jo05}. Note the evolution of the leading
    stream plane away from the trailing pole towards increasing $l$
    and the possible turn-over at $l_{\rm GC}\sim 280^{\circ}$ towards
    decreasing $b_{\rm GC}$. The amplitude of the plane precession is
    markedly different for Branches A and B. Also, note that the two
    branches have the opposing sense of the plane precession.
    \textit{Right:} Plane precession for both bright and faint Sgr
    trailing streams. The Gaussian scatter in distance and $B$ is
    added to the stream centroids of pairs of detections reported by
    \citet{Ko12}, each coloured point then gives the debris pole
    determined. There are 300 realisations for each of the 5 pairs of
    $\tilde{\Lambda}_{\odot}$. The colour changes in the fashion similar to
    that of the Left panel: from red to dark blue in the direction of
    the Sgr motion (also see Middle panel of
    Figure~\ref{fig:data_measure} for exact $\tilde{\Lambda}_{\odot}$
    values). While the precession of the bright trailing debris is
    consistent with that of the debris in Branch A of the leading
    tail, the debris in the faint trailing stream component precesses
    in the direction opposite to its bright counterpart.}
   \label{fig:pole}
\end{figure*}

The genesis of either the faint component of the Sgr stream or NGC
2419 is yet to be established. This paper shows that in 4 out of 6
phase-space coordinates, the stream and the cluster are coincident.
The recent discovery of highly peculiar elemental abundances in this
cluster by \citet{Co12} presents a rare opportunity to verify the
connection between the two. 

\section{The precession of the Sagittarius debris}
\subsection{Orbital precession}
\label{sec:precession}

The rate with which an orbit precesses in a spherically symmetric
gravitational potential depends primarily on how quickly the mass
generating the gravity field decays with radius. For example, in
Keplerian potentials, the precession angle is $0^{\circ}$ to ensure
the orbits are closed after one period, while in logarithmic haloes,
the precession is $\sim 120^{\circ}$. Orbits in the outer regions of
spherical galaxies should posses a precession rate somewhere between
$0^{\circ}$ and $\sim 120^{\circ}$. The precession rate is not solely
dependent on the mass decay rate in the host, it is also a weak
function of the orbital energy and angular momentum (see equation 3.18
in \citealt{BT}). The dependence of the orbital precession on the
potential and the eccentricity of the orbit is illustrated in
Figure~\ref{fig:precession_powerlaw}. The left panel of the Figure
shows rotation curves for the family of power law potentials $\Psi =
-\alpha^{-1} r^{-\alpha}$ for various $\alpha$. The right panel shows
the angle between the successive apo-centre passages in the potential
with given $\alpha$, for three orbital eccentricities. As can be seen
from Figure~\ref{fig:precession_powerlaw}, the precession angle
increases as the potential gets flatter. This increase is steeper for
more eccentric orbits.

It is not possible to measure the precession of the Sgr progenitor's
orbit directly, but the angle between the leading and the trailing
apo-centres encodes the necessary information. The angle between the
two apo-centres detected in the SDSS data is calculated as
follows. Two simple models are fitted to the Sgr Galactocentric
distance data: Gaussian and log-normal. While both are completely
unphysical, these describe the available data satisfactorily and give
an idea of the uncertainty associated with the model
mismatch. Overall, the distance to the trailing debris appears to be
more Gaussian-like, therefore the Gaussian model is fitted to all the
data presented in the top panel of
Figure~\ref{fig:data_measure}. Around the apo-centre, there is a clear
asymmetry in the run of the distances, hence the log-normal model is
fitted for the range $130^{\circ} < \tilde{\Lambda}_{\odot} <
200^{\circ}$. The opposite is true for the leading tail data: the
entire tail is modelled as log-normal, while Gaussian is fitted to the
distances in the range $35^{\circ} < \tilde{\Lambda}_{\odot} <
110^{\circ}$. For each of the tails, the average of the two model
values for the location of the apo-centre is taken.\footnote{Whenever
  the distance or the position angle formal error is smaller than the
  dispersion between the two models, the dispersion is taken as a
  proxy for the uncertainty.}

Solid lines of different colour show the best-fit models for both
tails in the middle panel of Figure~\ref{fig:data_measure}. For the
leading tail, the black line is the Gaussian and the blue curve is the
log-normal. The leading apo-centre is at $\tilde{\Lambda}_{\odot}^{\rm L} = 71
\fdg 3 \pm 3 \fdg 5$, where it reaches $R^{\rm L} = 47.8 \pm 0.5$
kpc. Trailing tail's apo-centre is at $\tilde{\Lambda}_{\odot}^{\rm T} = 170
\fdg 5 \pm 1^{\circ}$ and $R^{\rm T} = 102.5 \pm 2.5$ kpc. The red
curve shows the log-normal model, while black is the Gaussian. In the
heliocentric coordinate system of \citet{Ma03}, the differential
orbital precession between the leading and the trailing apo-centres is
$\delta \tilde{\Lambda}_{\odot} = 99 \fdg 3 \pm 3 \fdg 5$. Taking into account
the Sun's distance from the center of the Galaxy $R_0 = 8$ kpc, the
Galactocentric $\delta \tilde{\Lambda}_{\rm GC}= 93 \fdg 2 \pm 3 \fdg 5$.

Figure~\ref{fig:data_inplane} shows the Sgr stream detections
presented in Figure~\ref{fig:data_measure} now in the debris plane
defined by the pole at $(l_{\rm GC}, b_{\rm GC}) =
(275^{\circ},-14^{\circ})$. The right panel of the Figure compares the
current data with the model by \citet{LM10} projected onto the same
plane. While most of the data, including the entirety of the leading
tail detections and the trailing tail data in the range $220^{\circ} <
\tilde{\Lambda}_{\odot}< 280^{\circ}$ is reproduced by the model
extremely well, the distant trailing tail measurements in the range
$100^{\circ} < \tilde{\Lambda}_{\odot} < 200^{\circ}$ clearly do not
have any counterpart in this simulation. Both the distance to the
trailing apo-centre and its position angle are at odds with the
observations. Figure~\ref{fig:compare_vel} illustrates the model-data
mismatch further by displaying the run of the measured radial velocity
of both Sgr leading and trailing tidal tails as well as the evolution
of the LOS velocity centroid in the Sgr disruption model of
\citet{LM10}. While the model agrees perfectly with the kinematic data
at $220^{\circ} < \tilde{\Lambda}_{\odot} < 300^{\circ}$, there is a
small systematic offset at $80^{\circ} < \tilde{\Lambda}_{\odot} <
150^{\circ}$, but a pronounced gap at $150^{\circ} <
\tilde{\Lambda}_{\odot} < 190^{\circ}$. At first glance, it seems that
the new data presented here require the Galactic halo mass to drop
faster with increasing radius than imposed by the logarithmic model
used by \citet{LM10}.  The matter of inferring the Galactic mass
constraints from the Sgr debris detections presented here will be
addressed thoroughly in a separate publication (Gibbons et al, in
prep.)

\subsection{Orbital plane precession}
\label{sec:plane}

The previous Subsection assumed implicitly that the leading and the
trailing debris, at least up to their respective apo-centres, stay
within the same plane, namely the one defined by $(l_{\rm GC}, b_{\rm
  GC}) = (275^{\circ},-14^{\circ})$. Is this a reasonable assumption?
How fast does the plane of the debris precess? \citet{Jo05} show that
the M giants in the leading and the trailing tail define
Galactocentric planes whose poles are $\sim 10^{\circ}$ apart. Taking
advantage of the dramatic increase in the depth of the SDSS survey
compared to 2MASS, as well as the accurate stream distances based on
the BHB stars, it is timely to update the study of the evolution
of the plane of the Sgr debris.

Most of the new information supplied by the SDSS is in the North, and
so it is the leading tail plane evolution we will concentrate on. We
start by removing the Galactic foreground contribution from the
distribution of the MSTO star counts. This is done in the coordinate
system similar to the one used above, but rotated slightly to ensure
that both branches of the leading tail run at constant latitude
throughout the range of longitudes seen by the SDSS (see Left panel of
Figure~\ref{fig:data_pole}). Specifically, this is the plane defined
by the pole with $(l, b) = (99\fdg 7,13\fdg7)$. In each pixel of the
longitude $\Lambda_{\odot}^{\prime}$, the foreground density is fitted
with a linear model. Only the pixels outside the range $-8^{\circ} <
B_{\odot}^{\prime} < 10^{\circ}$ are taken into account during the
fit. The resulting model foreground is shown in the Middle panel of
Figure~\ref{fig:data_pole}. The residuals of the model subtraction are
shown in the Right panel of the Figure. With the foreground (dominated
by the Virgo stellar cloud) gone, the leading stream appears with
particular clarity.

Each pixel in the right panel of Figure~\ref{fig:data_pole} is a line
in the space of poles of Galactocentric planes, provided its distance
is known. The superposition of such lines defines the cloud of poles
corresponding to the group of pixels on the sky. Note that the pole
lines each carry a weight according to the stellar density of the
pixel they correspond to. We split Branches A and B into 5 overlapping
pieces with the boundaries marked in colour and shown below and above
each branch respectively, and map the stellar density in each of the
pieces onto the debris pole plane using the method above. Left panel
of Figure~\ref{fig:pole} gives the contours of the pole density for
each piece of Branch A and B. Pixels belonging to each branch are
selected using the simple condition of $B_{\odot}^{\prime} < 0$ for
Branch A, and $B_{\odot}^{\prime} > 0$ for Branch B. The colour-coding
used here is identical to the colour-coding in the Right panel of
Figure~\ref{fig:data_pole}. Also plotted here are the two measurements
based on the M giants of the trailing and the leading pole reported in
\citet{Jo05}, marked by filled square and triangle respectively. The
three lines crossing the figure are the lines of poles of the planes
passing through the current position of the remnant. The difference
between the lines is the assumed distance to the Sgr dSph.

The earliest it is possible to register the pole of the leading debris
is some $\sim 80^{\circ}$ away from the remnant. As shown by the red
contours, at this stage, the leading debris has already precessed away
several degrees from the plane defined by the M giants in the trailing
tail. The precession of the leading pole is in the direction of
increasing $l_{\rm GC}$ and decreasing $b_{\rm GC}$, the same
direction as identified by \citet{Jo05}. As we step along the Branch
A, the debris pole first moves further in the same direction, but then
turns before reaching the position determined using the M giants. It
is unclear whether this turn-around is real as most of the lines in
the space of poles contributed by pixels in the segments of the tail
at $\Lambda_{\odot}^{\prime} > 120^{\circ}$ are closely aligned as can
be gleaned from the shapes of the contours. The signature of the same
degeneracy can be seen in the top panels of Figure 3 of
\citet{Jo05}. Nonetheless, we can conclude with certainty that the
amplitude of the precession of the plane of the leading debris over
$\sim 100^{\circ}$ along the stream ought to be less than $\sim
10^{\circ}$.

The evolution of the debris plane defined by the pixels in Branch B is
surprisingly different from that of Branch A. The density contours are
even more stretched, making it difficult to pinpoint the exact
position of the pole for each section of the stream. However, the
broad-brush behaviour is apparent: the sense of precession of the
Branch B debris pole is opposite to that of Branch A. As one steps
along the leading tail away from the remnant in the direction of Sgr
motion, the Branch B pole moves in the direction of {\it decreasing}
$l$ and decreasing $b$.

In the South, we do not perform the full debris mapping into the space
of the orbital poles, as the SDSS coverage of the stream is
limited. Instead, we compute the poles of the planes defined by the
pairs of the first 6 detections given in Table 2 of
\citet{Ko12} \footnote{Beyond that, the stream position on the sky is
  uncertain due to incomplete coverage}. For each such pair, 300
Monte-Carlo trials are carried out to propagate the distance and the
latitude $\tilde{B}_{\odot}$ uncertainty. The resulting debris poles are shown
as elongated clouds of dots, whose colour-coding is explained in the
Middle panel of Figure~\ref{fig:data_measure}. The colour changes from
red to light blue in the direction of the Sgr motion, similar to the
colour-coding scheme used for the leading tail study above. It is
reassuring to see that the mean of the five debris pole clouds
generated using the bright component of the trailing stream shown in
the Figure lies very close to the value of \citet{Jo05}. Overall, the
poles of the debris in the bright trailing stream at $240^{\circ} <
\tilde{\Lambda}_{\odot} < 280^{\circ}$ have the same sense of precession as
the Branch A debris around $80^{\circ} < \tilde{\Lambda}_{\odot} <
100^{\circ}$, i.e. the stream angular momentum vector moves in the
direction of increasing $l$ and decreasing $b$. Once again, the faint
component of trailing stream shows the sense of precession opposite to
the bright portion of the tail: the pole moves in the direction of
decreasing $l$ and increasing $b$.

Finally, the poles of the debris planes at the positions of the
apo-centres seem to lie very close to each other, as judged by the
Branch A red contours in the Left panel and dark blue and light blue
dots (for the Bright arm) in the Right panel, and reassuringly close
to the pole of $275^{\circ}, -14^{\circ}$ chosen for orbit mapping in
Section~\ref{sec:precession}.

\begin{table*}
\caption{Heliocentric distances to the Sgr Leading stream based on the BHB detections.}
\begin{tabular}{|c c c c c c c c c c c c c c c c|}
\hline
$\Lambda_{\odot}$& 37.1 &  41.8 &  46.5 &  55.9 &  60.6 &  65.4 &  70.1 &  74.8 &  79.5 &  84.2 &  88.9 &  93.6 &  98.4 & 107.8 & 117.2 \\
$(m-M)_0$&17.83 & 18.12 & 18.27 & 18.83 & 18.65 & 18.66 & 18.56 & 18.51 & 18.44 & 18.47 & 18.41 & 18.25 & 17.96 & 17.73 & 17.30 \\
$\sigma_-$&0.20 & 0.05 & 0.04 & 0.23 & 0.06 & 0.03 & 0.03 & 0.03 & 0.06 & 0.05 & 0.07 & 0.06 & 0.14 & 0.11 & 0.09 \\
$\sigma_+$&0.06 & 0.06 & 0.05 & 0.15 & 0.05 & 0.03 & 0.03 & 0.03 & 0.04 & 0.08 & 0.08 & 0.05 & 0.20 & 0.11 & 0.11 \\
\hline
\end{tabular}
\label{tab:lead_dist}
\end{table*}

\begin{table*}
\caption{Heliocentric distances to the Sgr Trailing stream based on the BHB detections.}
\begin{tabular}{|c c c c c c c c c c|}
\hline
$\Lambda_{\odot}$&138.9 & 145.5 & 152.1 & 158.7 & 165.3 & 171.9 & 178.5 & 185.1 & 191.7 \\
$(m-M)_0$&19.11 & 19.54 & 19.49 & 19.76 & 19.85 & 19.92 & 19.98 & 19.74 & 19.49 \\
$\sigma_-$&0.06 & 0.07 & 0.07 & 0.03 & 0.05 & 0.03 & 0.07 & 0.03 & 0.03 \\
$\sigma_+$&0.11 & 0.15 & 0.05 & 0.03 & 0.03 & 0.04 & 0.07 & 0.05 & 0.03 \\
\hline
\end{tabular}
\label{tab:trail_dist}
\end{table*}

\begin{table*}
\caption{Galactocentric velocities of the Sgr Leading stream based on the giant star detections.}
\begin{tabular}{|c c c c c c c c c c c c c c c c|}
\hline
$\Lambda_{\odot}$& 62.7 &  68.8 &  75.0 &  81.2 &  87.3 &  93.5 &  99.6 & 105.8 & 111.9 & 118.1 & 124.2 & 130.4 & 136.5 & 142.7 & 148.8 \\
$V_{\rm GSR}$&  41.1 &    8.8 &   -7.6 &  -16.7 &  -29.5 &  -60.2 &  -70.7 &  -81.3 &  -90.8 & -102.7 & -109.2 & -111.6 & -118.3 & -121.7 & -115.5 \\
$\sigma_-$&10.4 &  3.4 &  5.1 &  6.4 &  5.9 &  4.4 &  3.9 &  2.9 &  2.1 &  5.0 &  2.4 &  2.4 &  1.7 &  3.4 &  7.9 \\
$\sigma_+$& 7.9 &  3.0 &  5.8 &  7.0 &  6.0 &  1.8 &  4.9 &  3.5 &  2.2 &  3.4 &  2.7 &  2.6 &  2.8 &  2.9 &  6.1 \\
\hline
\end{tabular}
\label{tab:lead_vel}
\end{table*}

\begin{table*}
\caption{Galactocentric velocities of the Northern Sgr Trailing stream based on the giant star detections.}
\begin{tabular}{|c c c c c c c c c|}
\hline
$\Lambda_{\odot}$&130.1 & 137.2 & 144.2 & 151.2 & 158.3 & 165.3 & 172.4 & 179.4 \\
$V_{\rm GSR}$& 127.4 &  132.5 &  128.8 &   77.0 &   44.7 &   17.8 &  -13.8 &  -31.3 \\
$\sigma_-$&17.1 &  3.6 &  2.5 &  4.4 &  1.5 &  7.9 & 17.0 & 22.4 \\
$\sigma_+$& 9.3 &  3.6 &  2.1 &  2.7 &  2.5 &  3.1 & 16.6 & 11.2 \\
\hline
\end{tabular}
\label{tab:trail_vel}
\end{table*}

\begin{table*}
\caption{Galactocentric velocities of the Southern Sgr Trailing stream based on the giant star detections.}
\begin{tabular}{|c c c c c c c c c c c c c c c|}
\hline
$\Lambda_{\odot}$&217.5 & 227.5 & 232.5 & 237.5 & 242.5 & 247.5 & 252.5 & 257.5 & 262.5 & 267.5 & 272.5 & 277.5 & 285.0 & 292.5 \\
$V_{\rm GSR}$&-127.2 & -141.1 & -150.8 & -141.9 & -135.1 & -129.5 & -120.0 & -108.8 &  -98.6 &  -87.2 &  -71.8 &  -58.8 &  -35.4 &   -7.8 \\
$\sigma_-$& 3.1 &  2.7 &  3.6 &  1.8 &  1.6 &  2.1 &  1.2 &  1.4 &  1.2 &  1.2 &  1.3 &  2.1 &  1.6 &  2.8 \\
$\sigma_+$& 3.1 &  2.7 &  3.6 &  1.8 &  1.6 &  2.1 &  1.2 &  1.4 &  1.2 &  1.2 &  1.3 &  2.1 &  1.6 &  2.8 \\
\hline
\end{tabular}
\label{tab:trail_vel_south}
\end{table*}

\section{Discussion and Conclusions}

\label{sec:disc}

We have taken a Swiss Army Knife approach to the SDSS database and
mapped the Sgr stellar stream across the sky using three stellar
tracers. The MSTO stars helped us define the plane of the Sgr debris,
the BHBs provided accurate distances, and the giants (mostly K and M)
gave the clearest view to date of the kinematics of both the leading
and the trailing tails as far as 100 kpc in the Galactic halo. Here,
is the summary of our main conclusions.

\medskip
\noindent
(1) As we push the SDSS data to its limits, that is to the edge of the
survey footprint, we are able to discern the apo-centres of the
leading and the trailing streams. Distant Sgr debris reported
previously by \citet{Ne03}, \citet{Ru11} and \citet{Be06} link with
the new BHB detections to reveal the wide arc of the trailing tail.

\medskip
\noindent
(2) Both the orbital precession of the trailing debris and the
Galactocentric distances it reaches are at odds with the current
state-of-the-art models of Sgr disruption. The angle between the
leading and the trailing apo-centres is $\delta \tilde{\Lambda}_{\rm GC}= 93
\fdg 2 \pm 3 \fdg 5$, while their respective distances are $R^{\rm L}
= 47.8 \pm 0.5$ kpc $R^{\rm T} = 102.5 \pm 2.5$ kpc.

\medskip
\noindent
(3) We substantiate the detections of the tidal tails and their 3D
evolution with the measurements of the debris' line of sight
velocity. In particular, we show that around the apo-centres stream's
$V_{\rm GSR}$ passes through zero.

\medskip
\noindent
(4) We demonstrate that a highly pure sample of spectroscopic M giant
stars can be extracted from the SDSS data. Across the entire sky, but
more importantly in the range $140^{\circ} < \tilde{\Lambda}_{\odot} <
190^{\circ}$, these stars appear to provide a strong independent
verification of both the distance and the velocity signals uncovered
with the help of the BHBs and the RGBs.

\medskip
\noindent
(5) We show that in 4 out of 6 phase-space coordinates, the peculiar
globular cluster NGC 2419 and the Sgr trailing tail coincide. NGC 2419
lies very close to the trailing apo-centre, nearer to the edge of the
stream. Given its position and the chemical abundance, it is not
impossible that the globular is in fact related to the faint,
metal-poor companion of the trailing tail discovered by \citet{Ko12}.

\medskip
\noindent
(6) The plane of the debris delineated by the MSTO stars and anchored
with the BHB distances evolves slowly, but noticeably, in the Galactic
potential. We also detect a turn-over in the path of the angular
momentum of the leading tail.

\medskip
\noindent
(7) The sense of precession of the debris plane of both secondary
components to the stream, Branch B in the North and the faint trailing
tail in the South, appears to be opposite to the main, bright parts of
the stream.

\bigskip
\noindent

Here, we have measured the Galactocentric angle between the apocentres
of the leading and trailing tails of the Sgr stream and the difference
between their respective distances. The angle through which the orbit
turns from one apocentre to the next is largely controlled by the
radial profile of the potential. Our measured value is inconsistent
with a flat rotation curve, or logarithmic potential, and indicates
that the Milky Way's dark matter density falls off more quickly than
isothermal. We have also detected the precession of the orbital plane
of the Sgr stream.  This wallowing, or gentle nutation, of the orbital
plane occurs naturally in mildly triaxial potentials, for which
evidence exists on other grounds for the Milky Way (e.g.,
\citealt{De13}). In practice, the components of the Sgr progenitor most
likely possessed some internal rotation or spin (e.g.,
\citealt{Pe10}), which can couple to the torques provided by the
triaxiality. The complex behaviour of the precession of the plane of
the Sgr debris that we have measured is most likely caused by an
interplay between the rotational kinematics of the stripped material
and the torques exerted by the triaxial gravity field.

It is crucial to note, however, that without observing the
uninterrupted connection between the Southern and the Northern stellar
debris, it is impossible to say with 100\% certainty that the
structure visible in the SDSS data at $140^{\circ} <
\tilde{\Lambda}_{\odot} < 190^{\circ}$ is indeed the extension of the
Sgr trailing arm. A range of other, admittedly less likely (as
illustrated in Section 3), options can not be ruled out at the
moment. Amongst these are: the contamination by other, previously
unknown distant stellar halo sub-structure; a stream from a yet
undiscovered companion body, presumably infalling in-sync with the Sgr
dwarf. Encouragingly, as Figure \ref{fig:compare_vel} illustrates, the
future line-of-sight velocity measurements in the range $190^{\circ} <
\tilde{\Lambda}_{\odot} < 210^{\circ}$ will provide a helpful
discriminant: in this part of the sky, the current disruption models
and the predictions from our analysis start to diverge drastically.

Above all, this study has uncovered the stupendous scale of the Sgr
stream. The debris stretches over Galactocentric radii from roughly 20
kpc out to at least 100 kpc. This makes the Sgr stream the single most
powerful probe of the Milky Way's dark halo. It extends well beyond
the gas rotation curve, which gives out at 20 kpc. As the Sgr stream
also possesses abundant bright tracers like BHBs, accurate distances
and velocities are obtainable almost everywhere. Provided we can
decode the message in its runes, it offers momentous constraints on
the underlying gravity field at unprecedentedly remote distances. It
truly is a touchstone for studies of the very distant dark halo.

\section*{Acknowledgments}
The research leading to these results has received funding from the
European Research Council under the European Union's Seventh Framework
Programme (FP/2007-2013) / ERC Grant Agreement n. 308024. VB and MIW
acknowledge financial support from the Royal
Society. S.K. acknowledges financial support from the STFC and the
ERC. MG acknowledges financial support from the RS. EO was partially
supported by NSF grant AST0807498.

\appendix

\section{Coordinate transformation}
\label{sec:appendix}

Here, we provide the equations for converting Equatorial ($\alpha$,
$\delta$) to the Sgr stream coordinate system
({$\tilde{\Lambda}_{\odot}$}, $\tilde{B}_{\odot}$) and back. Note that
this transformation (but for the right-handed coordinate system
pointing in the direction opposite to the Sgr motion) is completely
determined by the Euler angles given by \citet{Ma03}. We remind the
reader that the ($\tilde{\Lambda}_{\odot}, \tilde{B}_{\odot}$)
coordinates defined here can be easily converted into
($\Lambda_{\odot},B_{\odot}$) of \citet{Ma03} with
$\Lambda_{\odot}=360^{\circ}-\tilde{\Lambda}_{\odot}$ and
$B_{\odot}=-\tilde{B}_{\odot}$.

\begin{eqnarray*}
\tilde{\Lambda}  = &{\rm atan2}(&-0.93595354 \cos(\alpha) \cos(\delta) -  0.31910658 \sin(\alpha) \cos(\delta)+0.14886895 \sin(\delta),\nonumber \\
&& 0.21215555 \cos(\alpha) \cos(\delta)-0.84846291\sin(\alpha) \cos(\delta) - 0.48487186 \sin(\delta)) \nonumber
\\
\tilde{B}  = & \arcsin( & 0.28103559 \cos(\alpha) \cos(\delta) -0.42223415\sin(\alpha) \cos(\delta)
+0.86182209 \sin(\delta))\nonumber\\
\alpha  = & {\rm atan2}( & -0.84846291 \cos(\tilde{\Lambda}) \cos(\tilde{B}) -  0.31910658
\sin(\tilde{\Lambda}) \cos(\tilde{B})-0.42223415\sin(\tilde{B}),\nonumber \\
&& 0.21215555 \cos(\tilde{\Lambda}) \cos(\tilde{B})- 0.93595354\sin(\tilde{\Lambda}) \cos(\tilde{B}) + 0.28103559 \sin(\tilde{B}))
\nonumber
\\
\delta  = &  \arcsin ( &-0.48487186 \cos(\tilde{\Lambda}) \cos(\tilde{B})+0.14886895\sin(\tilde{\Lambda}) \cos(\tilde{B})+0.86182209 \sin(\tilde{B}))\nonumber
\end{eqnarray*}

\noindent where $\tan({\rm atan2}(y,x))=y/x$

\label{lastpage}

\end{document}